\documentclass[final,5p,times,twocolumn]{elsarticle}

\usepackage{booktabs}
\usepackage{multirow}
\usepackage{hyperref}
\usepackage[table]{xcolor}
\usepackage{bbding}
\usepackage{makecell}
\usepackage{graphicx}
\usepackage{subcaption}
\usepackage{tabularx}
\usepackage{array}
\usepackage{wrapfig}
\usepackage{pifont}

\usepackage{placeins}
\usepackage{comment}

\usepackage[utf8]{inputenc}
\usepackage{amssymb}
\usepackage{amsmath}
\usepackage[T1]{fontenc}

\usepackage{ulem}
\usepackage{color}
\usepackage{float}
\usepackage{bm}

\definecolor{mygray}{gray}{.9}
\definecolor{annotation}{RGB}{0, 153, 0}
\definecolor{key_words}{RGB}{236, 0, 141}
\definecolor{RowColor}{rgb}{0.95, 0.95, 1}
\definecolor{cgray}{RGB}{220,220,220}
\definecolor{lightblue}{RGB}{163,199,235}
\definecolor{darkblue}{RGB}{0,76,153}
\definecolor{citegrey}{HTML}{75878a}
\definecolor{updatagreen}{RGB}{80,100,40}
\definecolor{updatagrey}{HTML}{686461}
\definecolor{citecolor}{HTML}{2980b9}
\definecolor{linkcolor}{HTML}{c0392b}
\definecolor{mygreen}{RGB}{80,100,40}

\begin{document}

\begin{frontmatter}

\title{Morphological Decoupling-Based Skeletal Classification for Clinical Assessment of Malocclusion} 

\author[1,2,3]{Zhichun Jin\fnref{fn1}}
\ead{jinzhichun@126.com}

\author[4]{Zhicheng He\fnref{fn1}}
\ead{herichard88@gmail.com}

\author[1,2,3]{Hao Xu}

\author[1,2,3]{Dongyang Li}

\author[1,2,3]{Lin Wang}

\author[4]{Hongliang Ren}

\author[4]{Long Bai\corref{cor1}}
\ead{b.long@link.cuhk.edu.hk}

\affiliation[1]{organization={The Affiliated Stomatological Hospital of Nanjing Medical University},
    city={Nanjing},
    country={China}}

\affiliation[2]{organization={State Key Laboratory Cultivation Base of Research, Prevention and Treatment for Oral Diseases, Nanjing Medical University},
    city={Nanjing},
    country={China}}

\affiliation[3]{organization={Jiangsu Province Engineering Research Center of Stomatological Translational Medicine},
    city={Nanjing},
    country={China}}

\affiliation[4]{organization={Department of Electronic Engineering, The Chinese University of Hong Kong},
    city={Hong Kong SAR},
    country={China}}
    
\fntext[fn1]{Equal Contribution.}            
\cortext[cor1]{Corresponding author.}


\begin{abstract}
Malocclusion skeletal grading is a fundamental task in orthodontics, critical for diagnosis and treatment planning. Traditionally, cone-beam computed tomography (CBCT) is used for visual measurement, and the reconstructed lateral cephalograms are handed over to expert dentists for diagnosis. However, manual review is time-consuming, labor-intensive, and subject to inter-operator variability. Therefore, an automatic CBCT-based system is needed for reliable malocclusion skeletal grading. In this case, we develop TeethGNN, a novel graph-based framework designed to combine CBCT image features with morphological information for accurate and efficient malocclusion grading. TeethGNN utilizes a decoupled learnable decoder to directly predict key morphological indicators from CBCT images, eliminating the need for manual measurements. These morphological features are then fused with image features using a graph neural network (GNN), which effectively models the relationships between the modalities. To further enhance robustness and calibration, we introduce a collaborative calibration strategy. This strategy combines multi-scale graph adversarial perturbation for explicit calibration and nonlinear topological graph calibration for implicit confidence adjustment. Extensive experiments and ablation studies on our collected clinical dataset demonstrate that our malocclusion measurement system achieves 77.08\% in accuracy and 89.61\% in AUC, outperforming the compared state-of-the-art methods. These results validate the effectiveness of graph-based multimodal fusion and collaborative calibration in improving malocclusion grading performance. Our system shows strong potential for advancing computer-aided orthodontic diagnosis, providing an accurate and reliable solution for vision-based clinical measurement and diagnosis.
\end{abstract}


\end{frontmatter}

\section{Introduction}
\label{sec:introduction}
The malocclusion skeletal grading is a widely used diagnostic framework used to categorize malocclusion cases into three classes (Class I, II, and III) based on the positional relationship between the maxilla and mandibula~\cite{campbell2021angle, nielsen2019comprehensive, rinchuse1989ambiguities,wastell1988orthodontic,kim2020web}. Malocclusion affects patients’ oral functions, facial aesthetics, and mental health, making early intervention and treatment essential for achieving better outcomes. Skeletal grading represents different malocclusion patterns and their severity. Accurate classification results are crucial for orthodontists to develop appropriate treatment plans. Traditionally, this classification relies on medical measurement systems with cone-beam computed tomography (CBCT), where expert orthodontists perform visual inspections and manual measurements on reconstructed lateral cephalograms~\cite{hardy2012prevalence}. While effective, these methods are time-consuming and prone to inter-operator variability, particularly in complex cases~\cite{gracea2024artificial}.

Recently, artificial intelligence (AI) algorithms have been increasingly integrated with vision-based measurement systems in healthcare~\cite{bai2023surgicalvqla,sellers2024human,zhao2025rethinking}. In the field of dentistry, numerous approaches based on convolutional neural networks (CNNs) and Transformers have been combined with X-ray- or CBCT-based medical measurement systems to assist clinicians in diagnosis~\cite{schneider2022benchmarking, chandrashekar2022collaborative, lee2018diagnosis}. Existing medical measurement algorithms typically train a feature extractor to extract visual features from images, use a classification layer to output logits, and extract the class with the highest probability. While this approach performs efficient representation learning on medical images and often achieves good results, the black-box nature of such methods leaves room for improvement. Specifically, in clinical scenarios, the malocclusion skeletal grading is often based on specific morphological indicators of the CBCT reconstructed lateral cephalograms, such as the angle between nasion-subspinale and nasion-supramental (ANB), and the angle between the mandibular plane and the Frankfort horizontal plane (MP-FH). These indicators are typically measured manually or through morphological methods, with the classification results determined accordingly. Inspired by this process, our network aims to follow a similar approach: first decoupling the original input image to extract morphological information, and then integrating this information with the original image to assist in downstream orthodontic diagnosis tasks.

In practical applications of our measurement algorithm, although we aim to effectively integrate morphological information with image data, we intend for the model’s input to consist solely of the original image. Otherwise, if the input still requires manually measured morphological information, it would impose a significant burden on medical practitioners. To address this, we first design a decoupling learnable decoder to model the ANB and MP-FH morphological indicators. With the input of the model remaining in the single CBCT reconstructed lateral cephalogram, the network shall directly predict ANB and MP-FH through these decoders. Subsequently, we explored which model architecture is best suited for multimodal information fusion between the morphological indicators and the image features in this context. Simple approaches, such as concatenation, summation, or attention-based fusion, have been widely studied for multimodal tasks. However, we believe these approaches are not well-suited to our scenario because the morphological features are simple numerical values, making it difficult to extract meaningful features and effectively fuse them with image features.

In this case, an effective measurement solution is to use the Graph Neural Network (GNN)~\cite{scarselli2008graph,diao2024graph}. GNN is a deep learning model designed for processing graph-structured data and can effectively model information about nodes, edges, and their connections~\cite{lei2023graph,velivckovic2017graph}. The core principle of GNN is based on a message-passing mechanism, where each node aggregates information from its neighbors and updates its representation by combining it with its own features. This process allows GNN to gradually capture both local and global features within the graph structure. Through multi-layer propagation, GNN can learn embeddings for nodes, edges, or the entire graph. In our case, GNNs shall represent information from different modalities (i.e., morphological information and image features) as distinct graph nodes and learn the node relationships. This approach ensures that morphological information, even when represented as simple numerical values, is not overlooked. It also enables effective fusion without requiring complex feature extraction. Therefore, we construct separate graph nodes for the decoupled morphological information and image representations. Through graph representation learning (GRL), we establish meaningful connections between graph nodes formed by the decoupled morphological information and image features, facilitating better interaction between modalities and improving fusion performance.

However, although graph-based measurement approaches can effectively model relationships between different modalities, the shallow architecture of GNNs typically limits their depth compared to traditional CNNs~\cite{bai2025multimodal,wang2022gcl,yuan2025recognizing}. Studies have shown that this relatively shallow architecture may lead to the model being uncalibrated. In addition, GNNs rely heavily on the quality of neighboring information and graph structure, as well as smaller training datasets and dynamic graph characteristics, which further exacerbate the mismatch between predicted confidence and true probability. These factors make the output probabilities of GNNs less aligned with true confidence, making calibration crucial in practical applications~\cite{kong2022flag,wang2021cagcn}. An uncalibrated model results in a mismatch between predicted probabilities and actual accuracy, undermining reliability and performance. The model's confidence can no longer serve as a trustworthy indicator of correctness, rendering probability outputs less meaningful. In tasks involving threshold-based decision-making, uncalibrated models may lead to improper threshold settings, increasing false-positive or false-negative rates. These issues reduce the robustness and applicability of the model, limiting its use in real-world scenarios.

In our vision-based medical measurement system, in addition to innovatively using a decoupling GNN to integrate morphological indicators and image representations for malocclusion skeletal grading, we further address the issue of GNN miscalibration described above. Specifically, we perform network calibration at different stages of the GNN, thereby enhancing the accuracy and robustness of our model in diagnosis. First, during the construction of graph nodes, we introduce gradient-based adversarial perturbations into the embedding of graph nodes. Without altering the topology of the graph, we inject multi-scale perturbations into the embedding space of the nodes. By employing an efficient adversarial training strategy that integrates perturbation generation with model optimization, we improve the robustness of the model and achieve explicit calibration. Subsequently, when generating the logits output of the GNN, we adjust the confidence of the nodes by leveraging the graph's topology and nonlinear transformations. This ensures that the confidence of neighboring nodes becomes more consistent, achieving implicit calibration and improving the overall performance.
Our contributions can be summarized as:
\begin{itemize}
    \item[--] We propose an intelligent CBCT-based measurement system for malocclusion skeletal grading, in which we develop TeethGNN, a decoupled graph-based framework by modeling oral morphological information and visual representations. This enables precise and efficient measurement and diagnosis progress, assisting dentists in the process of decision-making and treatment planning.
    \item[--] We apply explicit and implicit calibration techniques to the different stages of our TeethGNN. We inject adversarial perturbations as explicit multi-scale adversarial calibration while constructing graph nodes, and apply implicit graph nonlinear topological calibration during the prediction stage. The proposed collaborative graph calibration strategy leads to robust and accurate diagnosis predictions of our TeethGNN.
    \item[--] We conduct extensive comparative and ablation experiments. The results demonstrate that our intelligent measurement system achieves superior performance in malocclusion skeletal grading, consistently outperforming existing classification solutions. Furthermore, our calibration design effectively improves the performance. These findings highlight the strong potential of our CBCT-based measurement system in computer-aided diagnosis.
\end{itemize}

\section{Related Work}

\subsection{Algorithm-assisted Skeletal Grading for Malocclusion}

\textcolor{black}{Recent advancements in AI have considerably improved the efficiency and accuracy of oral measurement and malocclusion grading~\cite{park2021use,olivetti2025predict,kim2023orthognathic,han2025facial}.} Various types of data have been utilized for this purpose, including occlusal contact information and traditional classification methods such as skeletal classification. For instance, Yao et al. developed a device for measuring bite force integrated with a random forest model, achieving 87.83\% accuracy~\cite{yao2022automatic}. This approach offers a non-invasive diagnostic tool that reduces radiation exposure. Additionally, multi-modal frameworks combining CBCT, panoramic X-rays, and 3D dental arch models are showing promise in improving diagnostic accuracy and robustness~\cite{kim2020malocclusion, nayak2024study}, enabling a more comprehensive analysis using multiple data sources.

Deep learning models, particularly CNNs, have emerged as the cornerstone of automated systems for malocclusion classification~\cite{juneja2024application, sabri2023classification, zhang2023deep}. CNN algorithms trained on lateral cephalograms have achieved impressive performance, with landmark detection rates exceeding 98\%~\cite{jiao2024deep}. Other studies using cascaded CNNs on posteroanterior (PA) cephalograms have demonstrated point-to-point error as low as 1.26mm~\cite{han2024accuracy}, while CNNs have also been applied to optimize performance with other imaging modalities, such as PA cephalograms~\cite{takeda2021landmark} and lateral facial photographs~\cite{shimamura2024accuracy}. Furthermore, recent studies have incorporated graph-based approaches, such as GNNs, to model neighborhood relationships in medical imaging, including dentistry~\cite{paul2024systematic, mohammadi2024medical}. GNNs capture both spatial and relational information, improving classification systems by integrating local feature extraction with relational modeling~\cite{zhang2024concept}. \textcolor{black}{This hybrid approach enhances the accuracy and robustness of malocclusion classification systems, marking notable progress in AI applications within orthodontics.}

Despite advancements in AI, several limitations persist in CNN-based methods for malocclusion grading, including small sample sizes, inability to predict the direction of landmark changes, and reliance on 2D images like lateral cephalograms, which omit 3D information~\cite{koval2023occlusal}. Additionally, variability in landmark definitions and input data quality affect accuracy. Our work addresses these challenges by applying GNNs to skeletal grading, improving the capture of complex spatial relationships and enhancing the robustness, accuracy, and generalizability of AI systems for orthodontics, especially in cases where traditional methods fall short.

\subsection{Graph Representation Learning in Medical Scenarios}

Building upon the foundational advancements, GRL has demonstrated immense potential in biomedicine. In genomics, GRL employs graph-based modeling of gene interactions to predict gene functions, identify associations between genes and diseases, and improve protein function prediction in protein-protein interaction networks, facilitating the identification of drug targets critical in drug discovery efforts~\cite{li2022graph,zhou2022graph,jha2022prediction}.
Medical knowledge graphs (MKGs) represent another transformative application. Constructed from heterogeneous data sources, GNNs facilitate applications like drug interaction prediction and personalized treatment recommendations~\cite{wu2023medical, gao2023medical}. GNNs have also demonstrated utility in predicting adverse drug reactions by learning from large-scale MKGs~\cite{li2024research, patel2024adverse, zhang2021prediction}.
In the domain of medical imaging analysis, GNNs extend their capabilities beyond traditional Convolutional Neural Networks (CNNs) by modeling the intricate spatial and topological relationships inherent in medical images. By representing pixels or regions as nodes and their spatial relationships as edges, GNNs have achieved state-of-the-art results in tasks such as tumor  detection~\cite{lotfy2023robust, ravinder2023enhanced}. By integrating data from modalities such as CT, MRI, and PET scans, GNNs construct comprehensive representations that account for the complementary strengths of each modality, which has been particularly beneficial in early cancer detection and the assessment of neurodegenerative diseases~\cite{zhang2023graph}. The modeling of organ and tissue structures also uses anatomical graphs, where nodes represent anatomical regions or landmarks, and edges capture the relationships between them. This representation enables organ shape analysis and automated diagnosis in congenital heart defect scenarios~\cite{zhang2021graph}.

GNNs have been extensively explored in dental applications. In particular, TeethGNN employs GNNs to directly process non-Euclidean 3D mesh data, enhancing the efficiency and robustness of tooth segmentation against challenges such as malocclusions and scanning noise~\cite{zheng2022teethgnn}. Similarly, TEANet introduces a node-erasable adaptive graph specifically designed for tooth arrangement in orthodontic treatments, addressing complex cases like tooth extraction and overcrowded dentition~\cite{li4996981teanet}. Furthermore, comparative studies on panoramic X-rays have demonstrated that GNNs outperform traditional convolutional neural networks (CNNs) by capturing non-Euclidean spatial relationships, resulting in superior diagnostic accuracy for dental diseases~\cite{thumati2023comparative}. However, most GNN applications in orthodontics are limited, with fewer studies focusing on their use in skeletal classification.

\section{Methods}

\subsection{AI-assisted System}

\begin{figure*}[!t]
\centering
\hspace{-0.0cm} 
\includegraphics[width=0.7\linewidth]{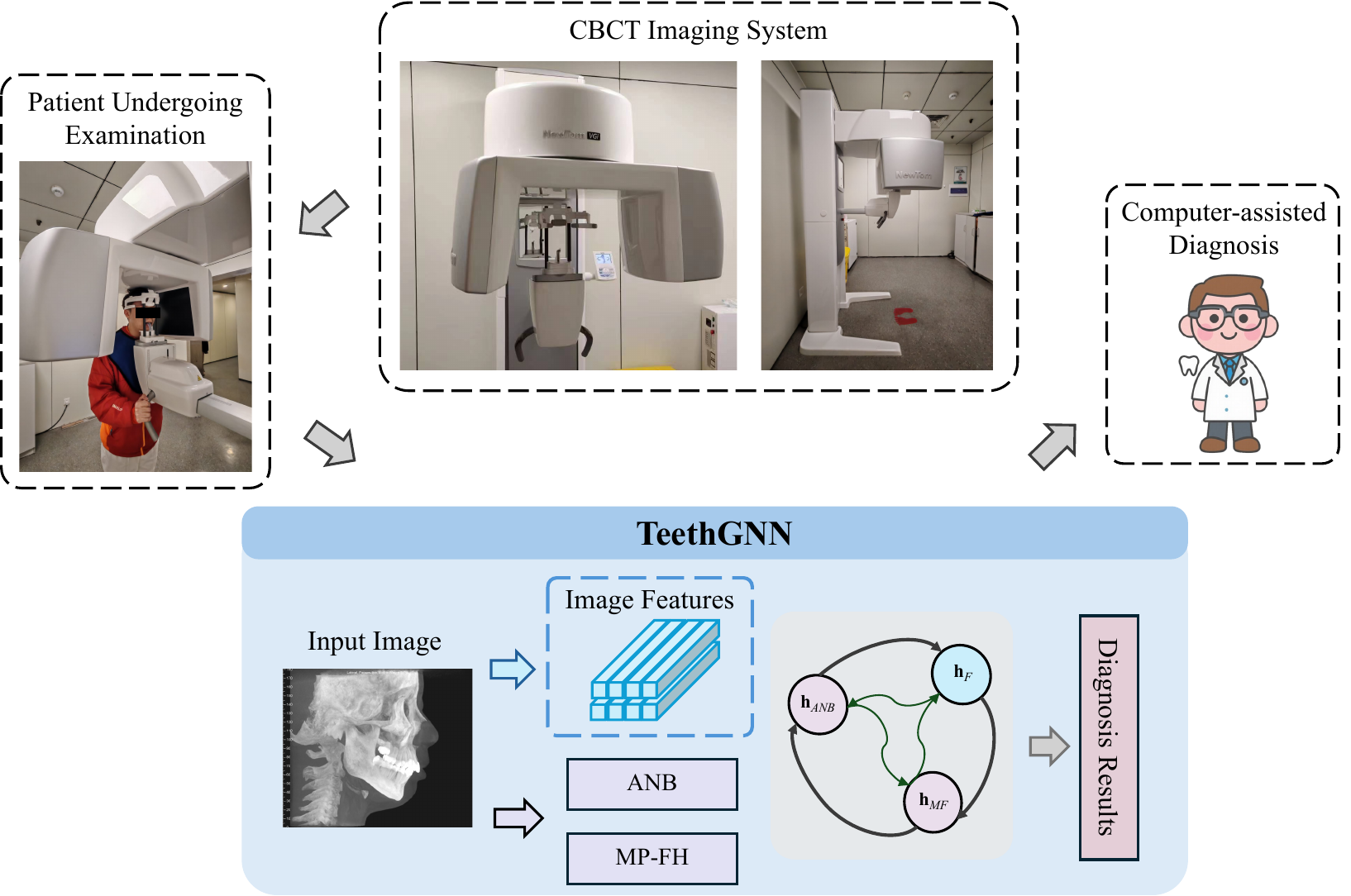}
\caption{Overview of our computer-assisted diagnosis system. TeethGNN will serve as the intelligent hub linking dentists and the imaging system.}
\label{fig:abs}
\end{figure*}

As shown in Fig.~\ref{fig:abs}, the proposed malocclusion skeletal grading system is based on the CBCT (NewTom VGi, Italy) imaging machine. This CBCT machine is equipped with a high-sensitivity flat-panel detector and a micro-focus X-ray tube, supporting various field-of-view (FOV) options to meet different clinical needs. For orthodontic evaluation, full cranial imaging was used with exposure parameters set to the FOV of 15 × 15 cm, a tube voltage of 110 kV, a tube current of 5–10 mA, and a focal spot size of 0.3 mm, enabling the reconstruction of lateral cephalograms. During the scan, the patient is positioned with the midsagittal plane parallel to the scanning plane, lips gently closed, and the upper and lower teeth in a natural occlusion position. Finally, the images are processed using our TeethGNN, and the prediction results are provided to dentists as diagnostic references.

\subsection{Preliminaries: GNNs}
Graph Neural Networks are designed to process graph-structured data. A graph is typically represented as $G = (V, E)$, where $V$ is the set of nodes and $E$ is the set of edges. For each node $v \in V$, a feature vector $\mathbf{x}_v$ is associated. The goal of GNNs is to learn meaningful node or graph representations by aggregating information from neighboring nodes. A common formulation of a GNN layer is:
\begin{equation}
  \mathbf{h}_v^{(k)} = \sigma \left( \mathbf{W}^{(k)} \cdot \text{AGG} \left( \left\{ \mathbf{h}_u^{(k-1)} : u \in \mathcal{N}(v) \right\} \cup \left\{ \mathbf{h}_v^{(k-1)} \right\} \right) \right)
\end{equation}
where $\mathbf{h}_v^{(k)}$ is the node embedding of $v$ at the $k$-th layer, $\mathcal{N}(v)$ denotes the neighbors of $v$, $\text{AGG}(\cdot)$ represents the aggregation function, $\mathbf{W}^{(k)}$ is a trainable weight matrix, and $\sigma(\cdot)$ is a non-linear activation function. The node embeddings are initialized as $\mathbf{h}_v^{(0)} = \mathbf{x}_v$ using the input features. By stacking multiple GNN layers, the model captures higher-order neighborhood information, enabling effective representation learning for tasks such as node classification, link prediction, and graph classification.

\subsection{Proposed Framework: TeethGNN}

\subsubsection{Multi-modality Teeth Graph}
\begin{figure*}[!t]
\centering
\hspace{-0.0cm} 
\includegraphics[width=\linewidth]{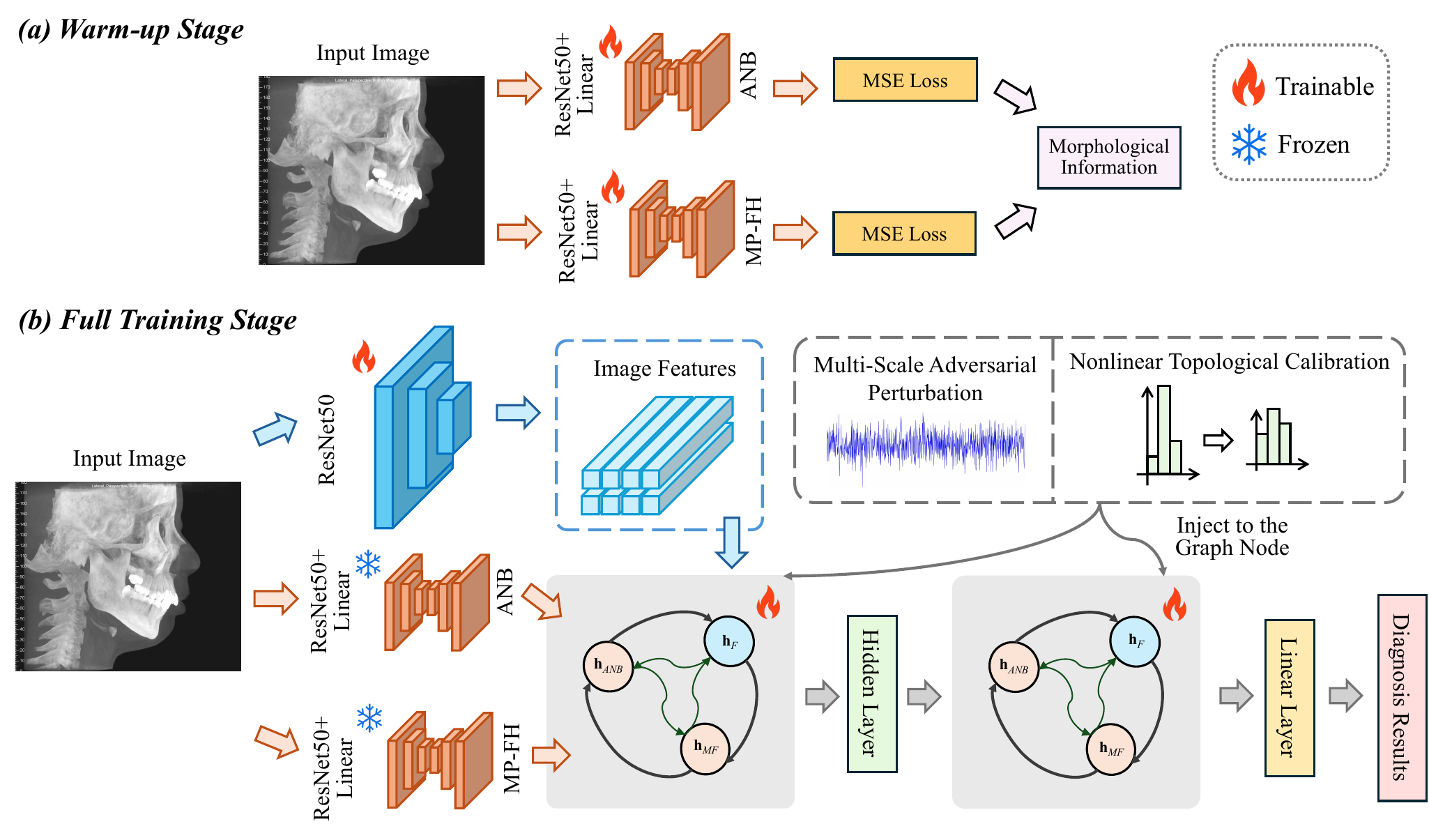}
\caption{
\textcolor{black}{Overview of the proposed framework. (a) Warm-up stage: the image encoder is trained using MSE loss to initialize morphological information extraction. (b) Full training stage: the extracted morphological information and image features are jointly fed into a graph network for representation learning. The explicit multi-scale adversarial calibration and implicit graph nonlinear topological calibration are applied to achieve accurate and robust diagnostic results. Trainable and frozen modules are marked separately.}}
\label{fig:main}
\end{figure*}

Our model architecture is illustrated in Fig.~\ref{fig:main}. TeethGNN processes the input image $I$ through two separate pathways: the image pathway and the morphological information pathway. These pathways are designed to capture complementary information: the image pathway focuses on extracting global visual features, and the morphological pathway learns specific morphological information critical for diagnosis.

In the image pathway, a standard ResNet50~\cite{he2016resnet} is employed as a feature extractor. ResNet50 is widely used for its ability to capture hierarchical features through residual connections, making it well-suited for visual representation learning. The extracted feature map $F$ can be formulated as:
\begin{equation}
F = \text{ResNet50}(I),
\end{equation}
where $F$ represents the high-dimensional feature representation of the input image $I$. These features provide a rich description of the image's global context, which is crucial for downstream diagnosis tasks.

In the morphological information pathway, we focus specifically on the morphological information ANB and MP-FH, which are essential for capturing the structural relationships within the image. To process these morphological features, we utilize a ResNet50~\cite{he2016resnet} as the encoder for each angular feature. The encoder is followed by a lightweight linear decoder to predict the respective angular values. This process can be described as:
\begin{equation}
\hat{y}_{\text{ANB}} = \text{Decoder}_{\text{ANB}}(\text{Encoder}_{\text{ANB}}(I)),
\end{equation}
\begin{equation}
\hat{y}_{\text{MP-FH}} = \text{Decoder}_{\text{MP-FH}}(\text{Encoder}_{\text{MP-FH}}(I)),
\end{equation}
where $\hat{y}_{\text{ANB}}$ and $\hat{y}_{\text{MP-FH}}$ denote the predicted angular values for ANB and MP-FH, respectively. Each encoder-decoder pair is trained to specialize in extracting and predicting the corresponding angular feature, ensuring accurate modeling of the morphological information.

To optimize the learning of these angular features, we adopt the Mean Squared Error (MSE) loss. The MSE loss is commonly used for regression tasks due to its simplicity and effectiveness in penalizing large deviations. The loss function for the morphological pathway is defined as:
\begin{equation}
\mathcal{L}_{\text{morph}} = \frac{1}{N} \sum_{i=1}^N \left( (y_{\text{ANB}}^{(i)} - \hat{y}_{\text{ANB}}^{(i)})^2 + (y_{\text{MP-FH}}^{(i)} - \hat{y}_{\text{MP-FH}}^{(i)})^2 \right),
\end{equation}
where $y_{\text{ANB}}^{(i)}$ and $y_{\text{MP-FH}}^{(i)}$ are the ground truth values for the $i$-th sample, and $N$ is the total number of training samples. This loss ensures that the predicted angular values closely match the ground truth, providing the model with precise morphological feature representations.

\textcolor{black}{After obtaining $\hat{y}_{\text{ANB}}$, $\hat{y}_{\text{MP-FH}}$, and $F$, we represent these complementary sources of information as a task-specific multimodal graph $G=(V,E)$, where $V=\{v_F, v_{\text{ANB}}, v_{\text{MP-FH}}\}$. The three nodes correspond to the image feature $F$, the predicted ANB measurement $\hat{y}_{\text{ANB}}$, and the predicted MP-FH measurement $\hat{y}_{\text{MP-FH}}$, respectively. A fully connected topology with self-loops is adopted so that each node can exchange information with both the image representation and the morphology-derived skeletal indicators. This compact topology provides an explicit fusion structure for the two scalar morphological measurements and the high-dimensional image feature, reducing the risk that the morphological information is diluted by direct feature concatenation. The GNN follows a graph-convolutional message-passing scheme, in which $\text{AGG}(\cdot)$ aggregates neighboring node embeddings and node-specific trainable transformations update each node representation. To learn the graph representations, the node embeddings in the GNN are iteratively updated as follows:}
\begin{equation}
\begin{aligned}
\mathbf{h}_F^{(k)} &= \sigma \left( \mathbf{W}_F^{(k)} \cdot \text{AGG} \left( \mathbf{h}_{\text{ANB}}^{(k-1)}, \mathbf{h}_{\text{MP-FH}}^{(k-1)}, \mathbf{h}_F^{(k-1)} \right) \right), \\
\mathbf{h}_{\text{ANB}}^{(k)} &= \sigma \left( \mathbf{W}_{\text{ANB}}^{(k)} \cdot \text{AGG} \left( \mathbf{h}_{\text{MP-FH}}^{(k-1)}, \mathbf{h}_F^{(k-1)}, \mathbf{h}_{\text{ANB}}^{(k-1)} \right) \right), \\
\mathbf{h}_{\text{MP-FH}}^{(k)} &= \sigma \left( \mathbf{W}_{\text{MP-FH}}^{(k)} \cdot \text{AGG} \left( \mathbf{h}_{\text{ANB}}^{(k-1)}, \mathbf{h}_F^{(k-1)}, \mathbf{h}_{\text{MP-FH}}^{(k-1)} \right) \right),
\end{aligned}
\end{equation}
where $\mathbf{h}_F^{(k)}$, $\mathbf{h}_{\text{ANB}}^{(k)}$, and $\mathbf{h}_{\text{MP-FH}}^{(k)}$ represent the embeddings of the image feature node, the ANB node, and the MP-FH node at the $k$-th layer, respectively. $\mathbf{W}_F^{(k)}$, $\mathbf{W}_{\text{ANB}}^{(k)}$, and $\mathbf{W}_{\text{MP-FH}}^{(k)}$ are trainable weight matrices associated with each node type at the $k$-th layer.
The initial node embeddings $\mathbf{h}_F^{(0)}$, $\mathbf{h}_{\text{ANB}}^{(0)}$, and $\mathbf{h}_{\text{MP-FH}}^{(0)}$ are initialized directly from their respective features, such that:
\begin{equation}
\mathbf{h}_F^{(0)} = F, \quad \mathbf{h}_{\text{ANB}}^{(0)} = \hat{y}_{\text{ANB}}, \quad \mathbf{h}_{\text{MP-FH}}^{(0)} = \hat{y}_{\text{MP-FH}}.
\end{equation}

By iteratively updating the node embeddings, the GNN effectively captures the interactions between the image feature and the morphological information, producing a unified representation of their relationships. After the GNN layers, the final embeddings of the nodes are concatenated into a single vector, which is passed through a linear layer for diagnosis classification. The diagnosis task is supervised using the cross-entropy loss, defined as:
\begin{equation}
\mathcal{L}_{\text{cls}} = - \frac{1}{N} \sum_{i=1}^N \sum_{c=1}^C y_i^c \log(\hat{y}_i^c),
\end{equation}
where $N$ is the total number of samples, $C$ is the number of classes, $y_i^c$ is the ground truth label for class $c$ of sample $i$, and $\hat{y}_i^c$ is the predicted probability for the same class. During the graph node construction and network inference, we further apply Multi-scale Graph Adversarial Perturbation and Nonlinear Topological Graph Calibration to help the model achieve more accurate and robust performance. The specific details are provided in Section~\ref{sec:pertubation} and Section~\ref{sec:topological}.

\subsubsection{Multi-scale Graph Adversarial Perturbation}
\label{sec:pertubation}

To improve the robustness and calibration of the GNN in malocclusion grading, we propose a multi-scale graph adversarial perturbation strategy inspired by~\cite{kong2022flag}, which introduces gradient-based perturbations into the graph node embeddings. Unlike methods that modify graph topology, our approach operates directly in the embedding space, preserving the graph’s structural integrity while enhancing feature diversity.
Given the constructed graph, we define the adversarial perturbation $\boldsymbol{\delta}$ as additive noise applied to the initial node embeddings. At the $t$-th iteration, the perturbed embeddings are formulated as:
\begin{equation}
\begin{aligned}
\mathbf{h}_F^{(0, t)} & = F + \boldsymbol{\delta}_F^{(t)}, \\
\mathbf{h}_{\text{ANB}}^{(0, t)} & = \hat{y}_{\text{ANB}} + \boldsymbol{\delta}_{\text{ANB}}^{(t)}, \\
\mathbf{h}_{\text{MP-FH}}^{(0, t)} & = \hat{y}_{\text{MP-FH}} + \boldsymbol{\delta}_{\text{MP-FH}}^{(t)},
\end{aligned}
\end{equation}
where $\boldsymbol{\delta}_F^{(t)}$, $\boldsymbol{\delta}_{\text{ANB}}^{(t)}$, and $\boldsymbol{\delta}_{\text{MP-FH}}^{(t)}$ are perturbations introduced to the corresponding node embeddings. These perturbations are optimized using a projected gradient ascent method. At each step, the perturbations are iteratively updated to maximize the training loss:
\begin{equation}
\boldsymbol{\delta}_v^{(t+1)} = \text{Proj}\left(\boldsymbol{\delta}_v^{(t)} + \lambda \cdot \text{sign}\left(\nabla_{\boldsymbol{\delta}_v^{(t)}} \mathcal{L}(\mathbf{h}_v^{(t)}, y)\right)\right),
\end{equation}
where $\boldsymbol{\delta}_v^{(t)}$ represents the perturbation for node $v$ at iteration $t$, $\lambda$ is the step size, and $\mathcal{L}$ is the training loss function. The projection operator $\text{Proj}(\cdot)$ ensures that the perturbations remain within a predefined norm ball to prevent excessive deviation from the original embeddings.

Additionally, we apply a multi-scale strategy by introducing perturbations with varying magnitudes. For each node type, the perturbations are scaled by factors $\{1, 2, \dots, S\}$ during training to create a diverse set of adversarial examples, which can be formulated as:
\begin{equation}
\boldsymbol{\delta}_v^{(t, m)} = s \cdot \lambda \cdot \text{sign}\left(\nabla_{\boldsymbol{\delta}_v^{(t)}} \mathcal{L}(\mathbf{h}_v^{(t)}, y)\right), \quad s \in \{1, 2, \dots, S\}.
\end{equation}
Therefore, this multi-scale approach improves the model’s ability to generalize by exposing it to a broad spectrum of input variations.

By efficiently integrating perturbation generation and model optimization, this approach minimizes additional computational costs while improving robustness. The multi-scale adversarial perturbations enhance the GNN's resilience to noisy inputs, and explicitly calibrate the network, ensuring accurate and reliable predictions for malocclusion grading.

\subsubsection{Nonlinear Topological Graph Calibration}
\label{sec:topological}
To enhance the trustworthiness and accuracy of malocclusion grading, we propose a nonlinear topological graph calibration mechanism. This module adjusts the output logits of the GNN by propagating confidence values across the graph topology, ensuring that neighboring nodes have more consistent confidence levels. By using nonlinear transformations, this approach implicitly calibrates the model while preserving its classification accuracy.

Given the uncalibrated logits of each node, denoted as $\mathbf{v}_{\text{ANB}}$, $\mathbf{v}_{\text{MP-FH}}$, and $\mathbf{v}_F$ for the morphological information $\hat{y}_{\text{ANB}}$, $\hat{y}_{\text{MP-FH}}$, and the image feature $F$, the calibration process applies a node-specific transformation:
\begin{equation}
\mathbf{v}_i' = \frac{\mathbf{v}_i}{t_i}, \quad \text{for } i \in \{F, \text{ANB}, \text{MP-FH}\},
\end{equation}
where $t_i > 0$ is a learned temperature parameter that scales the logits. This transformation ensures that the relative ranking of class probabilities remains unchanged, thus preserving the classification accuracy.

To further ensure consistency between the confidence values of connected nodes, we add a regularization term that minimizes the total variation of confidence across edges:
\begin{equation}
\mathcal{L}_{\text{cal}} = \sum_{(i, j) \in E} \left\| \mathbf{v}_i' - \mathbf{v}_j' \right\|_2^2,
\end{equation}
where $E$ represents the edges of the graph. This term encourages nodes with similar features or logits to have closer confidence levels after calibration.
This nonlinear calibration method ensures that the confidence values of nodes are consistent with the graph's structure, enhancing both the robustness and reliability of the GNN’s predictions.

\subsubsection{\textcolor{black}{Training and Inference Recipe}}

\textcolor{black}{To effectively train the model, we adopt a two-stage training strategy, including the warm-up and full training stages. We further describe the inference behavior to clarify the use of ground-truth and predicted morphological values.}
\\
\\
\noindent \textbf{Warm-up Training Stage.} In the warm-up stage, we exclusively train the morphological information pathway. This stage allows the model to focus solely on learning the morphological features $y_{\text{ANB}}$ and $y_{\text{MP-FH}}$ without interference from the classification task. By minimizing $\mathcal{L}_{\text{morph}}$ during this phase, the model is equipped with the ability to extract accurate angular features. The optimization function during the warm-up stage shall be represented as:
\begin{equation}
\mathcal{L}_{\text{warm}} = \mathcal{L}_{\text{morph}},
\end{equation}
This is the only phase in which the ground-truth angular values are used, namely as the regression targets of $\mathcal{L}_{\text{morph}}$.
\\
\\
\noindent \textbf{Full Training Stage.} In the full training stage, the morphological pathway is no longer supervised with ground-truth angular values. Instead, it operates as an auxiliary module to support the image pathway: for each image it produces the predicted angles $\hat{y}_{\text{ANB}}$ and $\hat{y}_{\text{MP-FH}}$, which are used to initialize the morphological graph nodes (Eq.~7). The training objective shifts to optimizing the diagnostic classification task, where the loss function is:
\begin{equation}
\mathcal{L}_{\text{full}} = \mathcal{L}_{\text{cls}} + \alpha \mathcal{L}_{\text{cal}},
\end{equation}
where $\alpha$ is a hyperparameter that balances the trade-off between classification accuracy and calibration smoothness. This stage ensures that the model leverages both global image features and auxiliary morphological information to improve its classification performance. Since the graph is built only from the predicted angles, our model uses only the image and diagnostic information as input and output, without including additional ground-truth morphological information, thereby eliminating any suspicion of `\textit{cheating}.' By combining the strengths of the image pathway and the morphological pathway, our model achieves a robust representation of both global and structural information, which is crucial for accurate diagnosis.
\\
\\
\noindent \textbf{Inference Stage.} During the inference stage, the behavior is consistent with the full training stage. The model takes only the lateral cephalogram as input, predicts $\hat{y}_{\text{ANB}}$ and $\hat{y}_{\text{MP-FH}}$ through the morphological pathway, initializes the morphological nodes with these predicted values, builds the graph, and outputs the skeletal class. No manual measurement is required, which matches the intended clinical use where ground-truth angular values are not available.

\section{Results}

\subsection{Dataset}

\begin{figure*}[!t]
\centering
\hspace{-0.0cm}
\includegraphics[width=\linewidth]{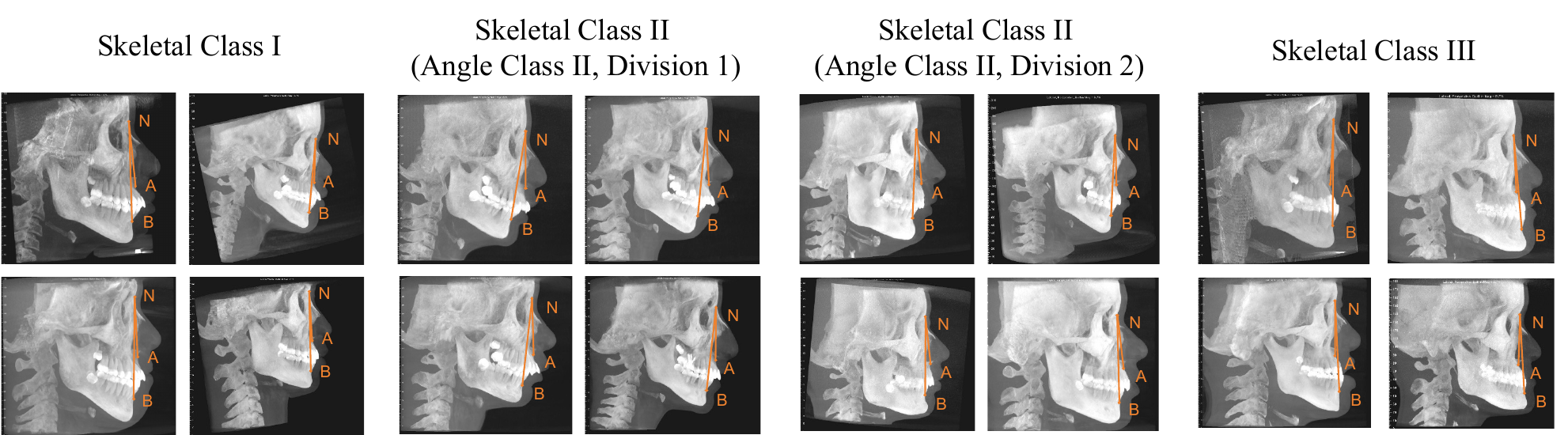}
\caption{Examples of four malocclusion categories. The orange lines show the position of the ANB angle.}
\label{fig:data}
\end{figure*}

The orthodontic data collection was conducted at the Anonymous Hospital, which contains CBCT-reconstructed lateral cephalograms from 476 patients undergoing malocclusion diagnosis. Specifically, the training and testing sets contain 428 and 48 samples, respectively, with cross-validation to make the results reliable. The dataset includes detailed coordinate annotations and cephalometric measurements such as ANB, MP-FH, etc. Expert dentists manually annotate the skeletal grading labels, which can be categorized into Skeletal Class I, Skeletal Class II (angle class II, division 1), Skeletal Class II (angle class II, division 2), and Skeletal Class III, as presented in Fig.~\ref{fig:data}. The study was approved by the Ethical Committee Department of the Anonymous Hospital.

\subsection{Implementation Details}
Our proposed solution is evaluated against the following comparison methods:
\newcounter{fnTorchvision}
\newcounter{fnTimm}
\newcounter{fnMMPretrain}
\newcounter{fnVMamba}
\newcounter{fnPyGCN}
\newcounter{fnGIN}
\newcounter{fnGAT}
\begin{itemize}
\setlength{\leftmargin}{0.1em}

    \item \textbf{CNN}: 
    ResNet18\footnote{https://github.com/pytorch/vision/tree/main}\setcounter{fnTorchvision}{\value{footnote}}~\cite{he2016resnet}, 
    ResNet50\footnotemark[\value{fnTorchvision}]~\cite{he2016resnet},
    RegNet\footnotemark[\value{fnTorchvision}]~\cite{xu2022regnet},
    EfficientNet\footnotemark[\value{fnTorchvision}]~\cite{tan2019efficientnet},
    MobilenetV2\footnotemark[\value{fnTorchvision}]~\cite{sandler2018mobilenetv2},
    ShuffleNet\footnotemark[\value{fnTorchvision}]~\cite{zhang2018shufflenet},
    DenseNet121\footnotemark[\value{fnTorchvision}]~\cite{huang2017densenet},
    ConvNeXtV2\footnote{https://github.com/huggingface/pytorch-image-models}\setcounter{fnTimm}{\value{footnote}}~\cite{woo2023convnext};

    \item \textbf{Transformer}: 
    ViT\footnotemark[\value{fnTimm}]~\cite{dosovitskiy2021vit}, 
    Swin\footnotemark[\value{fnTimm}]~\cite{liu2021swin}, 
    RIFormer\footnote{https://github.com/open-mmlab/mmpretrain}\setcounter{fnMMPretrain}{\value{footnote}}~\cite{wang2023riformer}, 
    MaxVit\footnotemark[\value{fnTimm}]~\cite{tu2022maxvit}, 
    HiViT\footnotemark[\value{fnMMPretrain}]~\cite{zhang2022hivit}, 
    EfficientFormer\footnotemark[\value{fnTimm}]~\cite{li2022efficientformer};

    \item \textbf{Foundation Models}: 
    DINOv2\footnotemark[\value{fnTorchvision}]~\cite{oquab2023dinov2}, 
    EVA-02\footnotemark[\value{fnTimm}]~\cite{fang2024eva};

    \item \textbf{Mamba}: 
    VMamba\footnote{https://github.com/MzeroMiko/VMamba}\setcounter{fnVMamba}{\value{footnote}}~\cite{liu2025vmamba};

    \item \textbf{GNN}: 
    VisionGNN\footnotemark[\value{fnMMPretrain}]~\cite{han2022visiongnn}, 
    GCN\footnote{https://github.com/tkipf/pygcn}\setcounter{fnPyGCN}{\value{footnote}}~\cite{kipf2016semi}, 
    GIN\footnote{https://github.com/weihua916/powerful-gnns}\setcounter{fnGIN}{\value{footnote}}~\cite{xu2018powerful}, 
    GAT\footnote{https://github.com/PetarV-/GAT}\setcounter{fnGAT}{\value{footnote}}~\cite{velivckovic2017graph}.

\end{itemize}

\textcolor{black}{All models were trained and evaluated under the same four-class label definition, training/test split, and test-time protocol. Data augmentation was applied only to the training images, while the test set is kept unchanged. The 12 augmentation operations were applied independently to the original images rather than compounded with one another, including resize, rotation, affine transformation, blur, color jitter, resized crop, grayscale conversion, horizontal flip, vertical flip, sharpness adjustment, random erase, and random perspective transformation.}

\textcolor{black}{TeethGNN was optimized with Adam using a learning rate of 0.0001, a batch size of 16, and up to 1000 epochs. The code for all baseline methods is sourced from the public repositories indicated in the footnotes. For image-only baselines (CNN, Transformer, Mamba, and GNN), the final classification heads were set as the four-class classifiers, and the models were fine-tuned end-to-end from official ImageNet-pretrained releases when such releases were publicly available. When official releases specified architecture-dependent input resolution, normalization, or fine-tuning settings, those settings were followed; when no specific fine-tuning recipe was provided, the common fine-tuning setting used Adam with a learning rate of 0.0001, a batch size of 16, and up to 1000 epochs. If ImageNet-pretrained weights were not publicly released for the exact variant, we used the official implementation or the released initialization provided by the model authors and trained the model under the same common schedule. For the foundation-model baselines, DINOv2 and EVA-02 were initialized from their official pretrained releases and fine-tuned with a four-class classifier. No hyperparameter was selected using the test labels.}

\subsection{Experimental Results}

\textcolor{black}{The experimental results in Table~\ref{tab:model_performance} demonstrate the performance advantages of our proposed TeethGNN model compared with the baseline methods across the major evaluation metrics.} \textcolor{black}{Among the CNN-based models, ResNet50 achieves the highest accuracy (0.7500), surpassing the other CNN models.} \textcolor{black}{However, even ResNet50 remains lower than TeethGNN: TeethGNN achieves an accuracy of 0.7708 and an F1-score of 0.6516, exceeding ResNet50 (0.7500 accuracy and 0.5600 F1-score) by gaps of 0.0208 and 0.0916, respectively.} This highlights the limitations of traditional convolutional architectures in fully capturing the structural and relational features inherent in our dataset.
\textcolor{black}{Transformer-based models generally perform worse than the best CNN baseline: ViT, MaxVit, and Swin reach accuracies of 0.5625, 0.6458, and 0.6667, respectively, all below ResNet50 (0.7500). This reflects the challenges of applying Transformers directly to medical imaging tasks without sufficient adaptation.} \textcolor{black}{Although Swin is the strongest Transformer baseline in accuracy (0.6667), it still falls short of TeethGNN (0.7708 accuracy and 0.6516 F1-score), with gaps of 0.1041 and 0.1529.} \textcolor{black}{Besides, foundation models show competitive results, particularly EVA-02, with an accuracy of 0.7083 and an F1-score of 0.5415. However, these models still fall below TeethGNN, which is higher by 0.0625 in accuracy and 0.1101 in F1-score, suggesting that pre-trained large-scale models may not be fully optimized for the specific needs of malocclusion grading.}

\begin{table*}[!t]
\caption{Comparison of experimental results against existing and state-of-the-art methods.}
\centering
\setlength{\tabcolsep}{22pt}  
\resizebox{\textwidth}{!}{
\begin{tabular}{lcccccc}
\toprule
\textbf{Models} & \textbf{Accuracy} & \textbf{Precision} & \textbf{Recall} & \textbf{F1-Score} & \textbf{AUC} \\
\midrule
ResNet18~\cite{he2016resnet} & 0.6667 & 0.4542 & 0.4103 & 0.4311 & 0.7949 \\
ResNet50~\cite{he2016resnet} & 0.7500 & 0.5506 & 0.5697 & 0.5600 & 0.8681 \\
MobilenetV2~\cite{sandler2018mobilenetv2} & 0.6250 & 0.4194 & 0.4325 & 0.4258 & 0.8016 \\
EfficientNet~\cite{tan2019efficientnet} & 0.6458 & 0.4520 & 0.4333 & 0.4425 & 0.8108 \\
RegNet~\cite{xu2022regnet} & 0.6667 & 0.4542 & 0.4103 & 0.4311 & 0.7460 \\
DenseNet121~\cite{huang2017densenet} & 0.6875 & 0.4821 & 0.4533 & 0.4673 & 0.7803 \\
ShuffleNet~\cite{zhang2018shufflenet} & 0.6875 & 0.5783 & 0.5451 & 0.5612 & 0.7799 \\
ConvNeXtV2~\cite{woo2023convnext} & 0.5625 & 0.4988 & 0.4501 & 0.4732 & 0.6183 \\
ViT~\cite{dosovitskiy2021vit} & 0.5625 & 0.3900 & 0.4098 & 0.3997 & 0.6986 \\
Swin~\cite{liu2021swin} & 0.6667 & 0.5790 & 0.4380 & 0.4987 & 0.7894 \\
MaxVit~\cite{tu2022maxvit} & 0.6458 & 0.5088 & 0.3984 & 0.4469 & 0.7079 \\
HiViT~\cite{zhang2022hivit} & 0.5000 & 0.3163 & 0.3284 & 0.3222 & 0.5134 \\
RIFormer~\cite{wang2023riformer} & 0.7292 & 0.5417 & 0.4845 & 0.5115 & 0.7082 \\
EfficientFormer~\cite{li2022efficientformer} & 0.6875 & 0.4617 & 0.4533 & 0.4575 & 0.7854 \\
EVA-02   (Finetuned)~\cite{fang2024eva} & 0.7083 & 0.5931 & 0.4982 & 0.5415 & 0.7223 \\
DINOv2 (Finetuned)~\cite{oquab2023dinov2} & 0.5625 & 0.4740 & 0.4112 & 0.4403 & 0.6871 \\
VMamba~\cite{liu2025vmamba} & 0.7083 & 0.3946 & 0.3803 & 0.3873 & 0.6847 \\
VisionGNN~\cite{han2022visiongnn} & 0.6667 & 0.4548 & 0.4453 & 0.4500 & 0.7385 \\
GAT~\cite{velivckovic2017graph} & 0.6042 & 0.5595 & 0.5948 & 0.5766 & 0.7521 \\
GCN~\cite{kipf2016semi} & 0.6875 & 0.6670 & 0.6097 & 0.6371 & 0.8287 \\
GIN~\cite{xu2018powerful} & 0.6667 & 0.5944 & 0.6260 & 0.6098 & 0.7791 \\
TeethGNN (\textbf{Ours}) & \textbf{0.7708} & \textbf{0.6778} & \textbf{0.6273} & \textbf{0.6516} & \textbf{0.8961} \\
\bottomrule
\end{tabular}}
\label{tab:model_performance}
\end{table*}

\textcolor{black}{When compared with VisionGNN, the only other graph-based model in the comparison, TeethGNN shows notable improvements.} \textcolor{black}{TeethGNN achieves an accuracy that is higher by 0.1041 and an F1-score that is higher by 0.2016 than VisionGNN (0.6667 accuracy and 0.4500 F1-score).} \textcolor{black}{This performance gap highlights the effectiveness of our graph-based design in leveraging structural information, which is critical for understanding the relationships in lateral cephalograms.} Furthermore, TeethGNN also achieves the highest AUC value (0.8961), indicating its superior ability to distinguish between classes with high confidence.
\textcolor{black}{In summary, TeethGNN achieves the best overall results, supporting its ability to address the challenges of malocclusion skeletal grading.} These results emphasize the importance of incorporating graph-based representations and customized modeling strategies for this complex orthodontic diagnosis task.

\subsection{Ablation Studies}

\begin{table*}[h]
\caption{Ablation study on the calibration components. We (i) remove the Multi-scale Graph Adversarial Perturbation (MGAP), and (ii) remove the Nonlinear Topological Graph Calibration (NTGC). \textcolor{black}{ECE and Brier score are reported as calibration metrics.}}
\centering
\resizebox{0.8\textwidth}{!}{
\begin{tabular}{cc|ccccccc}
\toprule
\textbf{MGAP} & \textbf{NTGC} & \textbf{Accuracy} & \textbf{Precision} & \textbf{Recall} & \textbf{F1-Score} & \textbf{AUC} & \textcolor{black}{\textbf{ECE}} & \textcolor{black}{\textbf{\makecell[c]{Brier\\ Score}}} \\
\midrule
$\times$ & $\times$ & 0.7500 & 0.6051 & 0.5743 & 0.5892 & 0.8653 & \textcolor{black}{0.1051} & \textcolor{black}{0.4910} \\
$\checkmark$ & $\times$ & 0.7292 & 0.6375 & 0.5010 & 0.5611 & 0.8916 & \textcolor{black}{0.2123} & \textcolor{black}{0.5098} \\
$\times$ & $\checkmark$ & 0.7708 & 0.6368 & \textbf{0.6365} & 0.6366 & 0.8732 & \textcolor{black}{0.2058} & \textcolor{black}{0.5080} \\
\textbf{$\checkmark$} & \textbf{$\checkmark$} & \textbf{0.7708} & \textbf{0.6778} & 0.6273 & \textbf{0.6516} & \textbf{0.8961} & \textcolor{black}{\textbf{0.0910}} & \textcolor{black}{\textbf{0.4987}} \\
\bottomrule
\end{tabular}}
\label{tab:components}
\end{table*}

\begin{table}[h]
\centering
\caption{Ablation study on the loss ratio $\alpha$.}
\resizebox{0.48\textwidth}{!}{
\begin{tabular}{cccccccc}
\toprule
\textbf{MODEL} & \textbf{Accuracy} & \textbf{Precision} & \textbf{Recall} & \textbf{F1-Score} & \textbf{AUC} \\
\midrule
$\alpha=0.01$ & 0.7708 & \textbf{0.6778} & \textbf{0.6273} & \textbf{0.6516} & \textbf{0.8961} \\
$\alpha=0.05$ & \textbf{0.8125} & 0.6033 & 0.5944 & 0.5988 & 0.8843 \\
$\alpha=0.1$ & 0.7500 & 0.6839 & 0.5936 & 0.6356 & 0.8951 \\
$\alpha=0.5$ & 0.7917 & 0.7349 & 0.6300 & 0.6784 & 0.8786 \\
$\alpha=1$ & 0.7708 & 0.8178 & 0.6385 & 0.7171 & 0.8403 \\
\bottomrule
\end{tabular}}
\label{tab:loss}
\end{table}

\begin{table}[h]
\centering
\caption{Ablation study on the scale factor $s$ in the graph adversarial perturbation.}
\resizebox{0.48\textwidth}{!}{
\begin{tabular}{lccccc}
\toprule
\textbf{MODEL} & \textbf{Accuracy} & \textbf{Precision} & \textbf{Recall} & \textbf{F1-Score} & \textbf{AUC} \\
\midrule
$s=1$ & 0.7292 & \textbf{0.7632} & 0.6000 & \textbf{0.6718} & 0.8472 \\
$s=2$ & 0.7500 & 0.7025 & 0.5771 & 0.6337 & 0.8841 \\
$s=3$ & \textbf{0.7708} & 0.6778 & \textbf{0.6273} & 0.6516 & \textbf{0.8961} \\
$s=4$ & \textbf{0.7708} & 0.5981 & 0.5559 & 0.5762 & 0.8847 \\
$s=5$ & 0.7292 & 0.6750 & 0.5651 & 0.6152 & 0.8744 \\
\bottomrule
\end{tabular}}
\label{tab:scale}
\end{table}

\begin{table}[htbp]
\centering
\caption{Ablation study on the fusion strategy. `Attn' denotes Attention.}
\resizebox{0.48\textwidth}{!}{
\begin{tabular}{lccccc}
\toprule
\textbf{FUSION} & \textbf{Accuracy} & \textbf{Precision} & \textbf{Recall} & \textbf{F1-Score} & \textbf{AUC} \\
\midrule
Concat~\cite{li2019visualbert} & 0.6250 & 0.5724 & 0.5964 & 0.5842 & 0.7279 \\
JCA~\cite{praveen2022joint} & 0.5833 & 0.4702 & 0.4817 & 0.4759 & 0.7097 \\
MAT~\cite{wu2022multimodal} & 0.5833 & 0.5491 & 0.4817 & 0.5132 & 0.7501 \\
Gated~\cite{arevalo2017gated} & 0.5625 & 0.5218 & 0.5651 & 0.5426 & 0.7077 \\
Self-Attn~\cite{vaswani2017attention} & 0.6458 & 0.5426 & 0.5203 & 0.5312 & 0.7565 \\
Guided-Attn~\cite{yu2019deep} & 0.5625 & 0.4722 & 0.4801 & 0.4761 & 0.7478 \\
Co-Attn (Bi)~\cite{yu2019deep} & 0.5625 & 0.4668 & 0.4621 & 0.4644 & 0.7153 \\
Co-Attn (V2T)~\cite{yu2019deep} & 0.5417 & 0.4631 & 0.4950 & 0.4785 & 0.7263 \\
Co-Attn (T2V)~\cite{yu2019deep} & 0.6042 & 0.4875 & 0.4297 & 0.4568 & 0.6238 \\
CAT-ViL~\cite{bai2023cat} & 0.5833 & 0.5249 & 0.5070 & 0.5158 & 0.7344 \\
\textbf{Ours} & \textbf{0.7708} & \textbf{0.6778} & \textbf{0.6273} & \textbf{0.6516} & \textbf{0.8961} \\
\bottomrule
\end{tabular}}
\label{tab:fusion}
\end{table}

\noindent \textbf{Effects of the Calibration Components.} Table~\ref{tab:components} presents the results of the ablation study on the two proposed modules: Multi-scale Graph Adversarial Perturbation (MGAP) and Nonlinear Topological Graph Calibration (NTGC). \textcolor{black}{In addition to the classification metrics, we report Expected Calibration Error (ECE) and Brier score to assess confidence calibration. The complete model with both MGAP and NTGC achieves the lowest ECE (0.0910) and Brier score (0.4987), while retaining the best accuracy (0.7708), F1-score (0.6516), and AUC (0.8961). These results suggest that the two calibration components improve confidence alignment, without compromising classification performance.}
\\
\\
\noindent \textbf{Effects of the Loss Ratio $\alpha$.} Table~\ref{tab:loss} shows the results of the ablation study on the loss ratio $\alpha$, which balances the cross-entropy loss and the calibration loss. When $\alpha = 0.05$, the model achieves the highest accuracy of 0.8125 and a competitive AUC of 0.8843, indicating that a small contribution from the calibration loss improves the model's overall performance. As $\alpha$ increases to 0.1 and 0.5, the performance slightly decreases in terms of accuracy, recall, and F1-score. However, the AUC increases to 0.8951 at $\alpha = 0.1$, suggesting that calibration loss may enhance the model's ability to distinguish classes at certain thresholds. Interestingly, when $\alpha = 1$, the model achieves the highest precision of 0.8178 and F1-score of 0.7171, but the accuracy drops to 0.7708 and the AUC reduces to 0.8403. This suggests that overly emphasizing calibration loss may lead to overfitting the calibration objective, reducing generalization. Overall, these results demonstrate the importance of tuning $\alpha$ to balance the trade-off between the primary classification task and the calibration objectives effectively. We ultimately selected $\alpha = 0.01$ as the final setting because it provides the most balanced performance across all metrics.
\\
\\
\noindent \textbf{Effects of the Scale Factor $s$.}
Table~\ref{tab:scale} presents the results of the ablation study on the scale factor $s$ in the multi-scale perturbation strategy. The scale factor $s$ determines how many levels of scale are included in the model. When $s = 1$, only scale 1 is used, leading to limited performance with an accuracy of 0.7292 and an AUC of 0.8472. Increasing $s$ to 2 results in moderate improvements, with an accuracy of 0.7500 and an AUC of 0.8841. The best performance is observed when $s = 3$, where scales 1, 2, and 3 are utilized. In this case, the model achieves the highest accuracy of 0.7708 and AUC of 0.8961, along with a recall of 0.6273, indicating a strong balance across the metrics. However, increasing $s$ beyond 3 does not further improve performance and even reduces precision and recall. These results highlight that using the 3-level scale strikes the optimal balance between the model's complexity and its performance.
\\
\\
\noindent \textbf{Effects of Different Fusion Strategies.}
\textcolor{black}{As shown in Table~\ref{tab:fusion}, compared to other multimodal fusion techniques, our proposed graph-based fusion gives the best results across these metrics.} First, our model showcases the top Accuracy of 0.7708, which is about 12\% higher than the best non-graph competitor (Self-Attention, 0.6458). \textcolor{black}{On the other balanced metrics, our model also performs well, with a precision of 0.6778 and a recall of 0.6273.} These results reveal that our model not only makes more correct predictions but also misses fewer true cases. The area under the ROC curve rises to 0.8961, giving an extra safety margin of roughly 0.14 over the strongest baseline. Such consistent improvement across five different measures shows that modeling the relations between modalities as a graph network helps the network capture complementary information and suppress noise that plain concatenation, attention, or gating cannot handle. In other words, the graph lets the model learn both local and global context, while still keeping the information flow simple and straightforward, which yields better generalization. \textcolor{black}{Therefore, our model maintains a good balance between precision and recall and produces more stable predictions than the other fusion schemes.}

\section{Discussion}

\noindent \textbf{Effectiveness and Clinical Relevance.}
\textcolor{black}{The experimental results show that our measurement solution outperforms the compared methods on most of the key evaluation metrics, highlighting the effectiveness of the proposed framework in malocclusion skeletal grading.} The model's ability to capture both global and structural relationships offers an improvement over conventional CNN-based and transformer-based approaches used in the field~\cite{bichu2025artificial, gracea2025artificial, sahim2025applications}. The incorporation of multi-scale adversarial perturbation and topological calibration further enhances robustness and generalizability. These advantages align with recent trends in multimodal orthodontic AI, such as the multiphase processing methods advocated by Bardideh et al.~\cite{bardideh2024designing}, reinforcing the value of combining diverse data sources for complex clinical tasks.
\textcolor{black}{Beyond these technical gains, the proposed system is also of practical clinical relevance. Designed as a decision-support tool rather than an autonomous diagnostic system, it predicts the skeletal class directly from the CBCT-reconstructed lateral cephalogram and can thus serve as a fast second reader that reduces manual measurement time and the inter-operator variability commonly seen in cephalometric analysis. Such a role is particularly useful in primary-care or community settings where orthodontic specialists are not always available, as the system can support early screening and timely referral, promote consistent diagnosis and treatment planning across operators, and aid the longitudinal monitoring of patients during treatment.}
\\
\\
\noindent \textbf{Limitations and Radiation Considerations.}
\textcolor{black}{Despite its promising performance, several limitations remain. First, the clinical cost of an error depends on which classes are confused. Confusions within the same skeletal type, such as Class~II Division~1 versus Division~2, share the same anteroposterior relationship and have a limited impact on the overall treatment direction. In contrast, confusions across types, especially those involving Class~III, are more consequential because they may correspond to different strategies such as growth modification, camouflage treatment, or orthognathic surgery. To mitigate this, the framework keeps a clinician in the loop, and the proposed calibration improves the reliability of the predicted confidence so that low-confidence or borderline cases can be flagged for expert review rather than acted upon directly.}

\textcolor{black}{Second, most errors arise in clinically ambiguous situations. These include borderline cases in which the morphological indicators lie close to the decision thresholds, such as an ANB value near the Class~I/Class~II or Class~II/Class~III boundary, the subtle distinction between Class~II Division~1 and Division~2 that relies on incisor inclination, and classes with fewer training samples. Notably, such cases are also frequent sources of disagreement among human experts. For this reason, TeethGNN is positioned to complement rather than replace specialist judgment. This view is consistent with broader AI-assisted orthodontic research, where Vaughan et al.~\cite{vaughan2025diagnostic} reported only fair agreement between an AI tool and clinicians in evaluating malocclusions, a result that reflects the intrinsic difficulty of these cases and the variability among experts themselves as much as the behavior of any single system. Accordingly, the system is intended to support clinical decision-making and to highlight cases that merit closer expert attention.}

\textcolor{black}{In addition, while TeethGNN leverages graph-based integration of image and morphological features, it has not yet incorporated dynamic functional parameters such as bite force or soft-tissue dynamics, which are important aspects of comprehensive diagnosis. Given the high variability and physiological complexity inherent in orthodontic conditions, integrating such multimodal data is itself challenging. As emphasized by Farook et al.~\cite{farook2025clinical}, the standardization of digital occlusion assessment remains difficult due to differences in expert philosophy and the multifactorial nature of occlusal dysfunction. Moreover, this study is based on a single-center dataset, which inevitably limits the diversity of imaging devices, acquisition protocols, and patient populations, so the cross-center generalizability of TeethGNN cannot yet be fully guaranteed. We therefore interpret our results as evidence of feasibility and internal validity on a single-center cohort rather than a claim of broad generalizability, and we regard external validation on independent cohorts as an essential next step toward clinical translation.}

\textcolor{black}{Lastly, radiation exposure is another important factor when applying CBCT-based systems in orthodontic diagnosis, and this concern is greater in growing patients and in cases that require repeated follow-up examinations, because the biological effects of ionizing radiation can accumulate over time. From this point of view, radiation-free 3D imaging is attractive, as it allows clinicians to perform more frequent and repeatable examinations with lower biological risk, which is useful for monitoring treatment progress and growth changes. Radiation-free 3D techniques, such as 3D facial scanning, have shown value for soft-tissue evaluation and longitudinal monitoring, and have been proposed as an alternative to CBCT volumetric rendering for soft-tissue assessment~\cite{perrotti2023radiation}. However, the present study targets skeletal malocclusion grading, which depends on craniofacial skeletal relationships and on morphological indicators such as ANB and MP-FH. These indicators are defined on bony landmarks and cannot yet be fully derived from facial surface scans, because surface scans capture soft-tissue shape rather than the underlying skeletal structure. For this reason, radiation-free 3D imaging cannot yet replace CBCT-derived skeletal information for our current task, although it remains promising for soft-tissue evaluation and follow-up. We therefore regard TeethGNN as a decision-support tool for cases in which CBCT is already clinically indicated, rather than as a reason to acquire additional radiation.}
\\
\\
\noindent \textbf{Future Works.}
\textcolor{black}{Future works will proceed along several directions. We will enlarge the dataset and improve its diversity and class balance to strengthen generalizability, particularly for rare classes and borderline cases. In particular, we will prioritize external validation on independent, multi-center cohorts spanning different institutions, CBCT devices, and patient populations, and explore federated learning to enable cross-institutional training and validation without centralizing patient data. We will also explore effective ways to embed additional modalities, such as biomechanical signals, soft-tissue dynamics, and electronic health records, into the graph learning framework to support more comprehensive diagnosis. In particular, we will combine the strengths of different imaging sources, so that CBCT can provide accurate skeletal information when it is clinically needed, while radiation-free 3D facial and intraoral scans can support soft-tissue analysis and frequent follow-up. Integrating CBCT with radiation-free 3D imaging within the same graph-based framework may reduce the dependence on ionizing radiation, allow more frequent and safer monitoring, and still maintain reliable diagnostic and prognostic performance. Finally, we plan to explicitly model boundary uncertainty and to strengthen human-in-the-loop mechanisms that account for the subjectivity and variability of clinical standards, so that the system can offer more reliable and clinically trustworthy decision support.}

\section{Conclusion}
This paper presents a novel vision-based medical measurement system for malocclusion skeletal grading. We develop TeethGNN, which integrates morphological information and CBCT image features using a graph-based approach. \textcolor{black}{The proposed framework incorporates two key modules: explicit multi-scale graph adversarial perturbation and implicit nonlinear topological graph calibration, which enhance model robustness, calibration, and diagnostic accuracy.} \textcolor{black}{Extensive experiments and ablation studies demonstrate the effectiveness of our system and its individual components, achieving strong overall performance compared with state-of-the-art methods.} Future work will focus on expanding the dataset to improve generalizability and exploring advanced graph-based strategies to further enhance the multimodal information integration. Additionally, efforts will be made to adapt the measurement system for broader applications in medical imaging and automatic diagnosis.

\section*{Author Contribution Statement}
\textbf{Zhichun Jin}: Conceptualization, Data curation, Formal analysis, Investigation, Methodology, Visualization, Writing – original draft. 
\textbf{Zhicheng He}: Conceptualization, Data curation, Formal analysis, Methodology, Software, Validation, Writing – original draft. 
\textbf{Hao Xu}: Data curation, Visualization, Software, Writing – review \& editing. 
\textbf{Dongyang Li}: Data curation, Visualization. Lin Wang: Conceptualization, Funding acquisition, Project administration, Resources, Supervision, Writing – review \& editing. 
\textbf{Hongliang Ren}: Conceptualization, Project administration, Resources, Supervision, Writing – review \& editing. 
\textbf{Long Bai}: Conceptualization, Investigation, Methodology, Project administration, Resources, Supervision, Writing – original draft.

\section*{Acknowledgment}
This work was supported by the Postgraduate Research \& Practice Innovation Program 597 of Jiangsu Province (KYCX22\_1801), and Natural Science Foundation of Jiangsu Province (BK20250581).

\appendix

\bibliographystyle{elsarticle-num-names}
\bibliography{reference}

@article{campbell2021angle,
  title={Angle's Classification--A Prosthodontic Consideration: Best Evidence Consensus Statement},
  author={Campbell, Stephen and Goldstein, Gary},
  journal={Journal of Prosthodontics},
  volume={30},
  number={S1},
  pages={67--71},
  year={2021},
  publisher={Wiley Online Library}
}

@article{nielsen2019comprehensive,
  title={A Comprehensive Diagnostic System for Orthodontists-Beyond Angle's Classification},
  author={Nielsen, Ib Leth},
  journal={Taiwanese Journal of Orthodontics},
  volume={31},
  number={3},
  pages={3},
  year={2019},
  publisher={Taiwan Association of Orthodontist}
}

@article{rinchuse1989ambiguities,
  title={Ambiguities of Angle's classification},
  author={Rinchuse, Donald J and Rinchuse, Daniel J},
  journal={The Angle Orthodontist},
  volume={59},
  number={4},
  pages={295--298},
  year={1989}
}

@article{wastell1988orthodontic,
  title={Orthodontic analysis and treatment planning: a suite of programs for performing centroid cephalometrics},
  author={Wastell, DG and Johnson, JS and Jones, JAH and Bennett, N},
  journal={Computer Methods and Programs in Biomedicine},
  volume={26},
  number={3},
  pages={259--265},
  year={1988},
  publisher={Elsevier}
}

@article{kim2020web,
  title={Web-based fully automated cephalometric analysis by deep learning},
  author={Kim, Hannah and Shim, Eungjune and Park, Jungeun and Kim, Yoon-Ji and Lee, Uilyong and Kim, Youngjun},
  journal={Computer methods and programs in biomedicine},
  volume={194},
  pages={105513},
  year={2020},
  publisher={Elsevier}
}

@article{hardy2012prevalence,
  title={Prevalence of angle class {III} malocclusion: A systematic review and meta-analysis},
  author={Hardy, Daniel K and Cubas, Yltze P and Orellana, Maria F},
  journal={Open Journal of Epidemiology},
  volume={2},
  number={4},
  pages={75--82},
  year={2012},
  publisher={Scientific Research Publishing},
  doi={10.4236/ojepi.2012.24012}
}

@article{gracea2024artificial,
  title={Artificial intelligence for orthodontic diagnosis and treatment planning: A scoping review},
  author={Gracea, Rellyca Sola and Winderickx, Nicolas and Vanheers, Michiel and Hendrickx, Julie and Preda, Flavia and Shujaat, Sohaib and de Llano Perula, Maria Cadenas and Jacobs, Reinhilde},
  journal={Journal of Dentistry},
  pages={105442},
  year={2024},
  publisher={Elsevier}
}

@INPROCEEDINGS{bai2023surgicalvqla,
  author={Bai, Long and Islam, Mobarakol and Seenivasan, Lalithkumar and Ren, Hongliang},
  booktitle={2023 IEEE International Conference on Robotics and Automation (ICRA)},
  title={Surgical-VQLA:Transformer with Gated Vision-Language Embedding for Visual Question Localized-Answering in Robotic Surgery},
  year={2023},
  pages={6859-6865}
}

@article{sellers2024human,
  title={Human autonomy teaming-based safety-aware navigation through bio-inspired and graph-based algorithms},
  author={Sellers, Timothy and Lei, Tingjun and Luo, Chaomin and Bi, Zhuming and Jan, Gene Eu},
  journal={Biomimetic Intelligence and Robotics},
  volume={4},
  number={4},
  pages={100189},
  year={2024},
  publisher={Elsevier}
}

@article{zhao2025rethinking,
  title={Rethinking data imbalance in class incremental surgical instrument segmentation},
  author={Zhao, Shifang and Bai, Long and Yuan, Kun and Li, Feng and Yu, Jieming and Dong, Wenzhen and Wang, Guankun and Hoque, Mobarak Islam and Padoy, Nicolas and Navab, Nassir and others},
  journal={Medical Image Analysis},
  volume={105},
  pages={103728},
  year={2025},
  publisher={Elsevier}
}

@article{schneider2022benchmarking,
  title={Benchmarking deep learning models for tooth structure segmentation},
  author={Schneider, L and Arsiwala-Scheppach, L and Krois, J and Meyer-L{\"u}ckel, Hendrik and Bressem, KK and Niehues, SM and Schwendicke, F},
  journal={Journal of dental research},
  volume={101},
  number={11},
  pages={1343--1349},
  year={2022},
  publisher={SAGE Publications Sage CA: Los Angeles, CA}
}

@article{chandrashekar2022collaborative,
  title={Collaborative deep learning model for tooth segmentation and identification using panoramic radiographs},
  author={Chandrashekar, Geetha and AlQarni, Saeed and Bumann, Erin Ealba and Lee, Yugyung},
  journal={Computers in Biology and Medicine},
  volume={148},
  pages={105829},
  year={2022},
  publisher={Elsevier}
}

@article{lee2018diagnosis,
  title={Diagnosis and prediction of periodontally compromised teeth using a deep learning-based convolutional neural network algorithm},
  author={Lee, Jae-Hong and Kim, Do-hyung and Jeong, Seong-Nyum and Choi, Seong-Ho},
  journal={Journal of periodontal \& implant science},
  volume={48},
  number={2},
  pages={114--123},
  year={2018},
  publisher={Korean Academy of Periodontology}
}

@article{scarselli2008graph,
  title={The graph neural network model},
  author={Scarselli, Franco and Gori, Marco and Tsoi, Ah Chung and Hagenbuchner, Markus and Monfardini, Gabriele},
  journal={IEEE transactions on neural networks},
  volume={20},
  number={1},
  pages={61--80},
  year={2008},
  publisher={IEEE}
}

@article{diao2024graph,
  title={Graph neural network based method for robot path planning},
  author={Diao, Xingrong and Chi, Wenzheng and Wang, Jiankun},
  journal={Biomimetic Intelligence and Robotics},
  volume={4},
  number={1},
  pages={100147},
  year={2024},
  publisher={Elsevier}
}

@article{lei2023graph,
  title={Graph-based robot optimal path planning with bio-inspired algorithms},
  author={Lei, Tingjun and Sellers, Timothy and Luo, Chaomin and Carruth, Daniel W and Bi, Zhuming},
  journal={Biomimetic Intelligence and Robotics},
  volume={3},
  number={3},
  pages={100119},
  year={2023},
  publisher={Elsevier}
}

@article{velivckovic2017graph,
  title={Graph attention networks},
  author={Veli{\v{c}}kovi{\'c}, Petar and Cucurull, Guillem and Casanova, Arantxa and Romero, Adriana and Lio, Pietro and Bengio, Yoshua},
  journal={arXiv preprint arXiv:1710.10903},
  year={2017}
}

@article{bai2025multimodal,
  title={Multimodal graph representation learning for robust surgical workflow recognition with adversarial feature disentanglement},
  author={Bai, Long and Ma, Boyi and Wang, Ruohan and Wang, Guankun and Cui, Beilei and Jiang, Zhongliang and Islam, Mobarakol and Min, Zhe and Lai, Jiewen and Navab, Nassir and others},
  journal={Information Fusion},
  volume={123},
  pages={103290},
  year={2025},
  publisher={Elsevier}
}

@inproceedings{wang2022gcl,
  title={Gcl: Graph calibration loss for trustworthy graph neural network},
  author={Wang, Min and Yang, Hao and Cheng, Qing},
  booktitle={Proceedings of the 30th ACM International Conference on Multimedia},
  pages={988--996},
  year={2022}
}

@inproceedings{yuan2025recognizing,
  title={Recognizing surgical phases anywhere: Few-shot test-time adaptation and task-graph guided refinement},
  author={Yuan, Kun and Chen, Tingxuan and Li, Shi and Lavanchy, Jo{\"e}l L and Heiliger, Christian and {\"O}zsoy, Ege and Huang, Yiming and Bai, Long and Navab, Nassir and Srivastav, Vinkle and others},
  booktitle={International Conference on Medical Image Computing and Computer-Assisted Intervention},
  pages={467--477},
  year={2025},
  organization={Springer}
}

@inproceedings{kong2022flag,
  title={Robust optimization as data augmentation for large-scale graphs},
  author={Kong, Kezhi and Li, Guohao and Ding, Mucong and Wu, Zuxuan and Zhu, Chen and Ghanem, Bernard and Taylor, Gavin and Goldstein, Tom},
  booktitle={Proceedings of the IEEE/CVF conference on computer vision and pattern recognition},
  pages={60--69},
  year={2022}
}

@article{wang2021cagcn,
  title={Be confident! towards trustworthy graph neural networks via confidence calibration},
  author={Wang, Xiao and Liu, Hongrui and Shi, Chuan and Yang, Cheng},
  journal={Advances in Neural Information Processing Systems},
  volume={34},
  pages={23768--23779},
  year={2021}
}

@inproceedings{park2021use,
  title={Use of artificial intelligence to predict outcomes of nonextraction treatment of Class II malocclusions},
  author={Park, Jae Hyun and Kim, Yoon-Ji and Kim, Jaehyun and Kim, Jinie and Kim, In-Hwan and Kim, Namkug and Vaid, Nikhilesh R and Kook, Yoon-Ah},
  booktitle={Seminars in Orthodontics},
  volume={27},
  number={2},
  pages={87--95},
  year={2021},
  organization={Elsevier}
}

@article{olivetti2025predict,
  title={How to predict the future face? A 3D methodology to forecast the aspect of patients after orthognathic surgeries},
  author={Olivetti, Elena Carlotta and Marcolin, Federica and Moos, Sandro and Vezzetti, Enrico and Borbon, Claudia and Zavattero, Emanuele and Ramieri, Guglielmo},
  journal={Computer Methods and Programs in Biomedicine},
  volume={265},
  pages={108757},
  year={2025},
  publisher={Elsevier}
}

@article{kim2023orthognathic,
  title={Orthognathic surgical planning using graph CNN with dual embedding module: External validations with multi-hospital datasets},
  author={Kim, In-Hwan and Kim, Jun-Sik and Jeong, Jiheon and Park, Jae-Woo and Park, Kanggil and Cho, Jin-Hyoung and Hong, Mihee and Kang, Kyung-Hwa and Kim, Minji and Kim, Su-Jung and others},
  journal={Computer Methods and Programs in Biomedicine},
  volume={242},
  pages={107853},
  year={2023},
  publisher={Elsevier}
}

@article{han2025facial,
  title={Facial surgery preview based on the orthognathic treatment prediction},
  author={Han, Huijun and Zhang, Congyi and Zhu, Lifeng and Singh, Pradeep and Hsung, Richard Tai-Chiu and Leung, Yiu Yan and Komura, Taku and Wang, Wenping and Gu, Min},
  journal={Computer Methods and Programs in Biomedicine},
  pages={108781},
  year={2025},
  publisher={Elsevier}
}

@inproceedings{yao2022automatic,
  title={Automatic Angle's classification based on the occlusal contact information},
  author={Yao, Zhiming and Wang, Peng and Li, Yan and Yang, Xianjun and Zhou, Xu and Wang, Yuanyin and Xu, Wenhua and Sun, Yining},
  booktitle={2022 IEEE International Conference on Systems, Man, and Cybernetics (SMC)},
  pages={761--767},
  year={2022},
  organization={IEEE}
}

@inproceedings{kim2020malocclusion,
  title={Malocclusion classification on 3D cone-beam CT craniofacial images using multi-channel deep learning models},
  author={Kim, Incheol and Misra, Dharitri and Rodriguez, Laritza and Gill, Michael and Liberton, Denise K and Almpani, Konstantinia and Lee, Janice S and Antani, Sameer},
  booktitle={2020 42nd Annual International Conference of the IEEE Engineering in Medicine \& Biology Society (EMBC)},
  pages={1294--1298},
  year={2020},
  organization={IEEE}
}

@inproceedings{nayak2024study,
  title={A Study on AI Applications for Orthodontics and Malocclusion Detection Approaches},
  author={Nayak, Ankitha A and Venugopala, PS and Ashwini, B and Padmashree, G},
  booktitle={2024 IEEE International Conference on Distributed Computing, VLSI, Electrical Circuits and Robotics (DISCOVER)},
  pages={273--278},
  year={2024},
  organization={IEEE}
}

@article{juneja2024application,
  title={Application of Convolutional Neural Networks for Dentistry Occlusion Classification},
  author={Juneja, Mamta and Saini, Sumindar Kaur and Kaur, Harleen and Jindal, Prashant},
  journal={Wireless Personal Communications},
  volume={136},
  number={3},
  pages={1749--1767},
  year={2024},
  publisher={Springer}
}

@inproceedings{sabri2023classification,
  title={Classification of Malocclusion using Convolutional Neural Network and Knowledge-Based Systems},
  author={Sabri, Fatin Anis Natasya Mohd and Ali, Azliza Mohd and Abd Rahman, Aida Nur Ashikin and Zurin, Mohd Amir Mukhsin and Salam, Afifah Syakirah Abdul and Din, Nur Amirah Che},
  booktitle={2023 IEEE 8th International Conference on Recent Advances and Innovations in Engineering (ICRAIE)},
  pages={1--4},
  year={2023},
  organization={IEEE}
}

@article{zhang2023deep,
  title={Deep learning-based prediction of mandibular growth trend in children with anterior crossbite using cephalometric radiographs},
  author={Zhang, Jia-Nan and Lu, Hai-Ping and Hou, Jia and Wang, Qiong and Yu, Feng-Yang and Zhong, Chong and Huang, Cheng-Yi and Chen, Si},
  journal={BMC Oral Health},
  volume={23},
  number={1},
  pages={28},
  year={2023},
  publisher={Springer}
}

@article{jiao2024deep,
  title={Deep learning for automatic detection of cephalometric landmarks on lateral cephalometric radiographs using the Mask Region-based Convolutional Neural Network: a pilot study},
  author={Jiao, Zhentao and Liang, Zhuangzhuang and Liao, Qian and Chen, Sheng and Yang, Hui and Hong, Guang and Gui, Haijun},
  journal={Oral Surgery, Oral Medicine, Oral Pathology and Oral Radiology},
  volume={137},
  number={5},
  pages={554--562},
  year={2024},
  publisher={Elsevier}
}

@article{han2024accuracy,
  title={Accuracy of posteroanterior cephalogram landmarks and measurements identification using a cascaded convolutional neural network algorithm: A multicenter study},
  author={Han, Sung-Hoon and Lim, Jisup and Kim, Jun-Sik and Cho, Jin-Hyoung and Hong, Mihee and Kim, Minji and Kim, Su-Jung and Kim, Yoon-Ji and Kim, Young Ho and Lim, Sung-Hoon and others},
  journal={Korean Journal of Orthodontics},
  volume={54},
  number={1},
  pages={48},
  year={2024},
  publisher={Korean Association of Orthodontists}
}

@article{takeda2021landmark,
  title={Landmark annotation and mandibular lateral deviation analysis of posteroanterior cephalograms using a convolutional neural network},
  author={Takeda, Saori and Mine, Yuichi and Yoshimi, Yuki and Ito, Shota and Tanimoto, Kotaro and Murayama, Takeshi},
  journal={Journal of Dental Sciences},
  volume={16},
  number={3},
  pages={957--963},
  year={2021},
  publisher={Elsevier}
}

@article{shimamura2024accuracy,
  title={Accuracy of cephalometric landmark and cephalometric analysis from lateral facial photograph by using CNN-based algorithm},
  author={Shimamura, Yui and Tachiki, Chie and Takahashi, Kaisei and Matsunaga, Satoru and Takaki, Takashi and Hagiwara, Masafumi and Nishii, Yasushi},
  journal={Scientific Reports},
  volume={14},
  number={1},
  pages={31089},
  year={2024},
  doi={10.1038/s41598-024-82230-z}
}

@article{paul2024systematic,
  title={A systematic review of graph neural network in healthcare-based applications: Recent advances, trends, and future directions},
  author={Paul, Showmick Guha and Saha, Arpa and Hasan, Md Zahid and Noori, Sheak Rashed Haider and Moustafa, Ahmed},
  journal={IEEE Access},
  volume={12},
  pages={15145--15170},
  year={2024},
  publisher={IEEE},
  doi={10.1109/ACCESS.2024.3354809}
}

@phdthesis{mohammadi2024medical,
  title={Medical Image Analysis Based on Graph Machine Learning and Variational Methods},
  author={Mohammadi, Sina},
  school={Chapman University},
  address={Orange, CA},
  year={2024},
  doi={10.36837/chapman.000601}
}

@article{zhang2024concept,
  title={The concept of AI-assisted self-monitoring for skeletal malocclusion},
  author={Zhang, Hexian and Liu, Chao and Yang, Pingzhu and Yang, Sen and Yu, Qing and Liu, Rui},
  journal={Health Informatics Journal},
  volume={30},
  number={3},
  pages={14604582241274511},
  year={2024},
  publisher={SAGE Publications Sage UK: London, England}
}

@article{koval2023occlusal,
  title={Occlusal characteristics in pre-orthodontic patients with deep overbite: An observational study},
  author={Koval, Svitlana and Sutter, Ben A and Koval, Stepan},
  journal={Advanced Dental Technologies \& Techniques},
  pages={1--8},
  year={2023},
  publisher={John Radke}
}

@article{li2022graph,
  title={Graph representation learning in biomedicine and healthcare},
  author={Li, Michelle M and Huang, Kexin and Zitnik, Marinka},
  journal={Nature Biomedical Engineering},
  volume={6},
  number={12},
  pages={1353--1369},
  year={2022},
  publisher={Nature Publishing Group UK London}
}

@article{zhou2022graph,
  title={Graph neural network for protein--protein interaction prediction: a comparative study},
  author={Zhou, Hang and Wang, Weikun and Jin, Jiayun and Zheng, Zengwei and Zhou, Binbin},
  journal={Molecules},
  volume={27},
  number={18},
  pages={6135},
  year={2022},
  publisher={MDPI}
}

@article{jha2022prediction,
  title={Prediction of protein--protein interaction using graph neural networks},
  author={Jha, Kanchan and Saha, Sriparna and Singh, Hiteshi},
  journal={Scientific Reports},
  volume={12},
  number={1},
  pages={8360},
  year={2022},
  publisher={Nature Publishing Group UK London}
}

@article{wu2023medical,
  title={Medical knowledge graph: Data sources, construction, reasoning, and applications},
  author={Wu, Xuehong and Duan, Junwen and Pan, Yi and Li, Min},
  journal={Big Data Mining and Analytics},
  volume={6},
  number={2},
  pages={201--217},
  year={2023},
  publisher={TUP}
}

@article{gao2023medical,
  title={Medical-knowledge-based graph neural network for medication combination prediction},
  author={Gao, Chao and Yin, Shu and Wang, Haiqiang and Wang, Zhen and Du, Zhanwei and Li, Xuelong},
  journal={IEEE Transactions on Neural Networks and Learning Systems},
  volume={35},
  number={10},
  pages={13246--13257},
  year={2024},
  publisher={IEEE},
  doi={10.1109/TNNLS.2023.3266490}
}

@inproceedings{li2024research,
  title={Research on adverse drug reaction prediction model combining knowledge graph embedding and deep learning},
  author={Li, Yufeng and Zhao, Wenchao and Dang, Bo and Yan, Xu and Gao, Min and Wang, Weimin and Xiao, Mingxuan},
  booktitle={2024 4th International Conference on Machine Learning and Intelligent Systems Engineering (MLISE)},
  pages={322--329},
  year={2024},
  organization={IEEE}
}

@inproceedings{patel2024adverse,
  title={Adverse Drug Reaction Prediction: Graph Neural Networks and Causal Inference Techniques},
  author={Patel, Jay and Patel, Rudra},
  booktitle={2024 4th Interdisciplinary Conference on Electrics and Computer (INTCEC)},
  pages={1--5},
  year={2024},
  organization={IEEE}
}

@article{zhang2021prediction,
  title={Prediction of adverse drug reactions based on knowledge graph embedding},
  author={Zhang, Fei and Sun, Bo and Diao, Xiaolin and Zhao, Wei and Shu, Ting},
  journal={BMC Medical Informatics and Decision Making},
  volume={21},
  pages={1--11},
  year={2021},
  publisher={Springer}
}

@article{lotfy2023robust,
  title={Robust tumor segmentation with hyperspectral imaging and graph neural networks},
  author={Lotfy, Mayar and Alperovich, Anna and Giannantonio, Tommaso and Barz, Bjorn and Zhang, Xiaohan and Holm, Felix and Navab, Nassir and Boehm, Felix and Schwamborn, Carolin and Hoffmann, Thomas K and others},
  journal={arXiv preprint arXiv:2311.11782},
  year={2023}
}

@article{ravinder2023enhanced,
  title={Enhanced brain tumor classification using graph convolutional neural network architecture},
  author={Ravinder, M and Saluja, Garima and Allabun, Sarah and Alqahtani, Mohammed S and Abbas, Mohamed and Othman, Manal and Soufiene, Ben Othman},
  journal={Scientific Reports},
  volume={13},
  number={1},
  pages={14938},
  year={2023},
  publisher={Nature Publishing Group UK London}
}

@article{zhang2023graph,
  title={Graph neural networks for image-guided disease diagnosis: A review},
  author={Zhang, Lin and Zhao, Yan and Che, Tongtong and Li, Shuyu and Wang, Xiuying},
  journal={iRADIOLOGY},
  volume={1},
  number={2},
  pages={151--166},
  year={2023},
  publisher={Wiley Online Library}
}

@article{zhang2021graph,
  title={Graph neural networks and their current applications in bioinformatics},
  author={Zhang, Xiao-Meng and Liang, Li and Liu, Lin and Tang, Ming-Jing},
  journal={Frontiers in genetics},
  volume={12},
  pages={690049},
  year={2021},
  publisher={Frontiers Media SA}
}

@article{zheng2022teethgnn,
  title={TeethGNN: semantic 3D teeth segmentation with graph neural networks},
  author={Zheng, Youyi and Chen, Beijia and Shen, Yuefan and Shen, Kaidi},
  journal={IEEE Transactions on Visualization and Computer Graphics},
  volume={29},
  number={7},
  pages={3158--3168},
  year={2022},
  publisher={IEEE}
}

@article{li4996981teanet,
  title={Teanet: Automated Tooth Extraction and Arrangement with Tooth-Level Graph Spatial Transformation Network},
  author={Li, Xiaoshuang and Chung, Miri and Bi, Lei and Huang, Jialiang and Pan, Yichen and Feng, David Dagan and Chen, Dong and Jiang, Lingyong and Sheng, Bin and Kim, Jinman},
  journal={Available at SSRN 4996981}
}

@inproceedings{thumati2023comparative,
  title={A comparative study on the working of gnn and cnn on panoramic x-rays in prediction of dental diseases},
  author={Thumati, Sai Manichandana Devi and Dhanya, Kode and Sathish, Harsha and Madan, KC Sekhar and Rani, Siji},
  booktitle={2023 8th International Conference on Communication and Electronics Systems (ICCES)},
  pages={755--762},
  year={2023},
  organization={IEEE}
}

@inproceedings{he2016resnet,
  title={Deep residual learning for image recognition},
  author={He, Kaiming and Zhang, Xiangyu and Ren, Shaoqing and Sun, Jian},
  booktitle={Proceedings of the IEEE conference on computer vision and pattern recognition},
  pages={770--778},
  year={2016}
}

@article{xu2022regnet,
  title={RegNet: Self-regulated network for image classification},
  author={Xu, Jing and Pan, Yu and Pan, Xinglin and Hoi, Steven and Yi, Zhang and Xu, Zenglin},
  journal={IEEE Transactions on Neural Networks and Learning Systems},
  volume={34},
  number={11},
  pages={9562--9567},
  year={2022},
  publisher={IEEE}
}

@inproceedings{tan2019efficientnet,
  title={Efficientnet: Rethinking model scaling for convolutional neural networks},
  author={Tan, Mingxing and Le, Quoc},
  booktitle={International conference on machine learning},
  pages={6105--6114},
  year={2019},
  organization={PMLR}
}

@inproceedings{sandler2018mobilenetv2,
  title={Mobilenetv2: Inverted residuals and linear bottlenecks},
  author={Sandler, Mark and Howard, Andrew and Zhu, Menglong and Zhmoginov, Andrey and Chen, Liang-Chieh},
  booktitle={Proceedings of the IEEE conference on computer vision and pattern recognition},
  pages={4510--4520},
  year={2018}
}

@inproceedings{zhang2018shufflenet,
  title={Shufflenet: An extremely efficient convolutional neural network for mobile devices},
  author={Zhang, Xiangyu and Zhou, Xinyu and Lin, Mengxiao and Sun, Jian},
  booktitle={Proceedings of the IEEE conference on computer vision and pattern recognition},
  pages={6848--6856},
  year={2018}
}

@inproceedings{huang2017densenet,
  title={Densely connected convolutional networks},
  author={Huang, Gao and Liu, Zhuang and Van Der Maaten, Laurens and Weinberger, Kilian Q},
  booktitle={Proceedings of the IEEE conference on computer vision and pattern recognition},
  pages={4700--4708},
  year={2017}
}

@inproceedings{woo2023convnext,
  title={Convnext v2: Co-designing and scaling convnets with masked autoencoders},
  author={Woo, Sanghyun and Debnath, Shoubhik and Hu, Ronghang and Chen, Xinlei and Liu, Zhuang and Kweon, In So and Xie, Saining},
  booktitle={Proceedings of the IEEE/CVF Conference on Computer Vision and Pattern Recognition},
  pages={16133--16142},
  year={2023}
}

@inproceedings{dosovitskiy2021vit,
  title={An Image is Worth 16x16 Words: Transformers for Image Recognition at Scale},
  author={Alexey Dosovitskiy and Lucas Beyer and Alexander Kolesnikov and Dirk Weissenborn and Xiaohua Zhai and Thomas Unterthiner and Mostafa Dehghani and Matthias Minderer and Georg Heigold and Sylvain Gelly and Jakob Uszkoreit and Neil Houlsby},
  booktitle={International Conference on Learning Representations},
  year={2021}
}

@inproceedings{liu2021swin,
  title={Swin transformer: Hierarchical vision transformer using shifted windows},
  author={Liu, Ze and Lin, Yutong and Cao, Yue and Hu, Han and Wei, Yixuan and Zhang, Zheng and Lin, Stephen and Guo, Baining},
  booktitle={Proceedings of the IEEE/CVF international conference on computer vision},
  pages={10012--10022},
  year={2021}
}

@article{wang2023riformer,
  title={Riformer: Keep your vision backbone effective while removing token mixer},
  author={Wang, Jiahao and Zhang, Songyang and Liu, Yong and Wu, Taiqiang and Yang, Yujiu and Liu, Xihui and Chen, Kai and Luo, Ping and Lin, Dahua},
  journal={arXiv preprint arXiv:2304.05659},
  year={2023}
}

@inproceedings{tu2022maxvit,
  title={Maxvit: Multi-axis vision transformer},
  author={Tu, Zhengzhong and Talebi, Hossein and Zhang, Han and Yang, Feng and Milanfar, Peyman and Bovik, Alan and Li, Yinxiao},
  booktitle={European conference on computer vision},
  pages={459--479},
  year={2022},
  organization={Springer}
}

@article{zhang2022hivit,
  title={Hivit: Hierarchical vision transformer meets masked image modeling},
  author={Zhang, Xiaosong and Tian, Yunjie and Huang, Wei and Ye, Qixiang and Dai, Qi and Xie, Lingxi and Tian, Qi},
  journal={arXiv preprint arXiv:2205.14949},
  year={2022}
}

@article{li2022efficientformer,
  title={Efficientformer: Vision transformers at mobilenet speed},
  author={Li, Yanyu and Yuan, Geng and Wen, Yang and Hu, Ju and Evangelidis, Georgios and Tulyakov, Sergey and Wang, Yanzhi and Ren, Jian},
  journal={Advances in Neural Information Processing Systems},
  volume={35},
  pages={12934--12949},
  year={2022}
}

@article{oquab2023dinov2,
  title={Dinov2: Learning robust visual features without supervision},
  author={Oquab, Maxime and Darcet, Timoth{\'e}e and Moutakanni, Th{\'e}o and Vo, Huy and Szafraniec, Marc and Khalidov, Vasil and Fernandez, Pierre and Haziza, Daniel and Massa, Francisco and El-Nouby, Alaaeldin and others},
  journal={arXiv preprint arXiv:2304.07193},
  year={2023}
}

@article{fang2024eva,
  title={Eva-02: A visual representation for neon genesis},
  author={Fang, Yuxin and Sun, Quan and Wang, Xinggang and Huang, Tiejun and Wang, Xinlong and Cao, Yue},
  journal={Image and Vision Computing},
  volume={149},
  pages={105171},
  year={2024},
  publisher={Elsevier}
}

@article{liu2025vmamba,
  title={Vmamba: Visual state space model},
  author={Liu, Yue and Tian, Yunjie and Zhao, Yuzhong and Yu, Hongtian and Xie, Lingxi and Wang, Yaowei and Ye, Qixiang and Jiao, Jianbin and Liu, Yunfan},
  journal={Advances in neural information processing systems},
  volume={37},
  pages={103031--103063},
  year={2025}
}

@article{han2022visiongnn,
  title={Vision gnn: An image is worth graph of nodes},
  author={Han, Kai and Wang, Yunhe and Guo, Jianyuan and Tang, Yehui and Wu, Enhua},
  journal={Advances in neural information processing systems},
  volume={35},
  pages={8291--8303},
  year={2022}
}

@article{kipf2016semi,
  title={Semi-supervised classification with graph convolutional networks},
  author={Kipf, Thomas N and Welling, Max},
  journal={arXiv preprint arXiv:1609.02907},
  year={2016}
}

@article{xu2018powerful,
  title={How powerful are graph neural networks?},
  author={Xu, Keyulu and Hu, Weihua and Leskovec, Jure and Jegelka, Stefanie},
  journal={arXiv preprint arXiv:1810.00826},
  year={2018}
}

@article{li2019visualbert,
  title={Visualbert: A simple and performant baseline for vision and language},
  author={Li, Liunian Harold and Yatskar, Mark and Yin, Da and Hsieh, Cho-Jui and Chang, Kai-Wei},
  journal={arXiv preprint arXiv:1908.03557},
  year={2019}
}

@inproceedings{praveen2022joint,
  title={A joint cross-attention model for audio-visual fusion in dimensional emotion recognition},
  author={Praveen, R Gnana and de Melo, Wheidima Carneiro and Ullah, Nasib and Aslam, Haseeb and Zeeshan, Osama and Denorme, Th{\'e}o and Pedersoli, Marco and Koerich, Alessandro L and Bacon, Simon and Cardinal, Patrick and others},
  booktitle={Proceedings of the IEEE/CVF conference on computer vision and pattern recognition},
  pages={2486--2495},
  year={2022}
}

@inproceedings{wu2022multimodal,
  title={Multimodal crowd counting with mutual attention transformers},
  author={Wu, Zhengtao and Liu, Lingbo and Zhang, Yang and Mao, Mingzhi and Lin, Liang and Li, Guanbin},
  booktitle={2022 IEEE International Conference on Multimedia and Expo (ICME)},
  pages={1--6},
  year={2022},
  organization={IEEE}
}

@article{arevalo2017gated,
  title={Gated multimodal units for information fusion},
  author={Arevalo, John and Solorio, Thamar and Montes-y-G{\'o}mez, Manuel and Gonz{\'a}lez, Fabio A},
  journal={arXiv preprint arXiv:1702.01992},
  year={2017}
}

@article{vaswani2017attention,
  title={Attention is all you need},
  author={Vaswani, Ashish and Shazeer, Noam and Parmar, Niki and Uszkoreit, Jakob and Jones, Llion and Gomez, Aidan N and Kaiser, {\L}ukasz and Polosukhin, Illia},
  journal={Advances in neural information processing systems},
  volume={30},
  pages={5998--6008},
  year={2017}
}

@inproceedings{yu2019deep,
  title={Deep modular co-attention networks for visual question answering},
  author={Yu, Zhou and Yu, Jun and Cui, Yuhao and Tao, Dacheng and Tian, Qi},
  booktitle={Proceedings of the IEEE/CVF conference on computer vision and pattern recognition},
  pages={6281--6290},
  year={2019}
}

@inproceedings{bai2023cat,
  title={Cat-vil: Co-attention gated vision-language embedding for visual question localized-answering in robotic surgery},
  author={Bai, Long and Islam, Mobarakol and Ren, Hongliang},
  booktitle={International Conference on Medical Image Computing and Computer-Assisted Intervention},
  pages={397--407},
  year={2023},
  organization={Springer}
}

@incollection{bichu2025artificial,
  title={Artificial Intelligence Applications in Orthodontics},
  author={Bichu, Yashodhan M and Zou, Bingshuang and Chaudhari, Prabhat Kumar and Adel, Samar M and Vaiid, Nikhillesh},
  booktitle={Artificial Intelligence for Oral Health Care: Applications and Future Prospects},
  pages={81--97},
  year={2025},
  publisher={Springer}
}

@article{gracea2025artificial,
  title={Artificial intelligence for orthodontic diagnosis and treatment planning: A scoping review},
  author={Gracea, Rellyca Sola and Winderickx, Nicolas and Vanheers, Michiel and Hendrickx, Julie and Preda, Flavia and Shujaat, Sohaib and de Llano-P{\'e}rula, Maria Cadenas and Jacobs, Reinhilde},
  journal={Journal of Dentistry},
  volume={152},
  pages={105442},
  year={2025},
  publisher={Elsevier}
}

@incollection{sahim2025applications,
  title={Applications of Artificial Intelligence and Machine Learning for Orthodontic Diagnosis},
  author={Sahim, Soukaina and Boutissante, Moncef and El Quars, Farid},
  booktitle={Cranio-Maxillofacial Surgery -- Orthognathic and Orthodontic Techniques},
  year={2025},
  publisher={IntechOpen},
  doi={10.5772/intechopen.1009333}
}

@article{bardideh2024designing,
  title={Designing an artificial intelligence system for dental occlusion classification using intraoral photographs: A comparative analysis between artificial intelligence-based and clinical diagnoses},
  author={Bardideh, Erfan and Alizadeh, Farzaneh Lal and Amiri, Maryam and Ghorbani, Mahsa},
  journal={American Journal of Orthodontics and Dentofacial Orthopedics},
  volume={166},
  number={2},
  pages={125--137},
  year={2024},
  publisher={Elsevier}
}

@article{vaughan2025diagnostic,
  title={Diagnostic accuracy of artificial intelligence for dental and occlusal parameters using standardized clinical photographs},
  author={Vaughan, Matthew and Mheissen, Samer and Cobourne, Martyn and Ahmed, Farooq},
  journal={American Journal of Orthodontics and Dentofacial Orthopedics},
  volume={167},
  number={6},
  pages={733--740},
  year={2025},
  publisher={Elsevier},
  doi={10.1016/j.ajodo.2025.01.017}
}

@article{farook2025clinical,
  title={Clinical machine learning in parafunctional and altered functional occlusion: A systematic review},
  author={Farook, Taseef Hasan and Rashid, Farah and Ahmed, Saif and Dudley, James},
  journal={The Journal of Prosthetic Dentistry},
  volume={133},
  number={1},
  pages={124--128},
  year={2025},
  publisher={Elsevier}
}

@article{perrotti2023radiation,
  title={A radiation free alternative to CBCT volumetric rendering for soft tissue evaluation},
  author={Perrotti, Giovanna and Reda, Rodolfo and Rossi, Ornella and D'Apolito, Isabella and Testori, Tiziano and Testarelli, Luca and others},
  journal={Brazilian Dental Science},
  volume={26},
  number={1},
  pages={e3726},
  year={2023},
  doi={10.4322/bds.2023.e3726}
}

\end{document}